\documentclass[useAMS,usegraphicx,usenatbib]{mn2e} 

\usepackage{ulem}
\usepackage{color}
\usepackage{graphicx}
\usepackage{times} 
\usepackage{amssymb}
\usepackage{amsmath}
\usepackage{lscape}
\usepackage{url}
\newif\ifAMStwofonts
\AMStwofontstrue

\def\rms{${\it r}_{\rm ms}$}
\def\rg{${\it r}_{\rm g}$}
\def\rin{${\it r}_{\rm in}$}
\def\laor{\rm{\sc LAOR}}
\def\phabs{\rm{\sc PHABS}}
\def\diskbb{\rm{\sc DISKBB}}
\def\refhiden{\rm{\sc REFHIDEN}}
\def\reflionx{\rm{\sc REFLIONX}}
\def\smedge{\rm{\sc SMEDGE}}
\def\kdblur{\rm{\sc KDBLUR}}
\def\pexriv{\rm{\sc PEXRIV}}
\def\nh{${\it N}_{\rm H}$}
\def\ka{$K\alpha$}

\def\epicpn{{\it EPIC}{\rm-pn}}
\def\epicmos1{{\it EPIC}{\rm-MOS1}}
\def\epicmos2{{\it EPIC}{\rm-MOS2}}
\def\epicmos{{\it EPIC}{\rm-MOS}}

\def\xmm{{\it XMM-Newton}}
\def\rxte{{\it RXTE}}
\def\hexte{{\it HEXTE}}
\def\xspec{\hbox{\sc XSPEC}}
\def\xspecv{{\sc XSPEC}{\rm\thinspace v\thinspace 11.3.2}}

\def\s{\hbox{$\rm\thinspace s$}}
\def\ks{\hbox{$\rm\thinspace ks$}}

\def\cm{\hbox{$\rm\thinspace cm$}}

\def\kpc{\hbox{$\rm\thinspace kpc$}}

\def\ev{\hbox{$\rm\thinspace eV$}}
\def\kev{\hbox{$\rm\thinspace keV$}}

\def\msun{\hbox{$\rm\thinspace M_{\odot}$}}

\def\gx{\hbox{\rm GX 339-4}}

\begin{document}

\title[] {A systematic look at the Very High and Low/Hard state of {GX\,339-4}: Constraining the black hole spin with a new reflection model} \author[R. C. Reis et al.]
{\parbox[]{6.in} {R.~C.~Reis $^{1}$\thanks{E-mail:
      rcr36@ast.cam.ac.uk}, A.~C.
    Fabian$^{1}$, R.~Ross$^{2}$, G.~Miniutti$^{1,5}$, J.~M. Miller$^{3}$ and C.~Reynolds$^{4}$\\ } \\
  \footnotesize
  $^{1}$Institute of Astronomy, Madingley Road, Cambridge, CB3 0HA\\ 
  $^{2}$Physics Department, College of the Holy Cross, Worcester, MA 01610, USA\\ 
  $^{3}$Department of Astronomy, University of Michigan, 500 Church Street, Ann Arbor, MI 48109, USA\\
  $^{4}$Department of Astronomy, The University of Maryland, College Park, MD, 20742, USA\\ 
  $^{5}$Laboratoire Astroparticule et Cosmologie (APC), UMR 7164, 10 Rue A. Domon et L. Duquet, 75205, Paris}

\maketitle

\begin{abstract}
  We present a systematic study of \gx\ in both its very high and low
  hard states from simultaneous observations made with \xmm\ and
  \rxte\ in 2002 and 2004. The X-ray spectra of both these
  extreme states exhibit strong reflection signatures, with a broad,
  skewed Fe-\ka\ line clearly visible above the continuum. Using a
  newly developed, self-consistent reflection model which implicitly
  includes the blackbody radiation of the disc as well as the effect
  of Comptonisation, blurred with a relativistic line function, we
  were able to infer the spin parameter of \gx\ to be $0.935 \pm 0.01$
  (statistical) $\pm \thinspace 0.01$ (systematic) at 90 per cent
  confidence. We find that both states are consistent with an ionised
  thin accretion disc extending to the innermost stable circular orbit
  around the rapidly spinning black hole.

\end{abstract}


\section{Introduction}

X-ray spectra of Galactic black hole candidates (GBHCs) are an
important tool in the studies of the inner regions of accretion flow
around black holes (BHs), providing information on both the geometry
of the accretion disc and on intrinsic physical parameters such as BH
mass and spin.

The spectrum can be explained by the combination of a quasi-thermal
blackbody component{\footnote{The term quasi-thermal blackbody applies
    here to the local emission at the surface of the accretion
    disc. Due to opacity effects it is broader than a true blackbody
    spectrum. } caused by radiatively efficient accretion through a
  disc (Shimura \& Takahara 1995; Merloni, Fabian \& Ross 2000), a
  power-law component due to inverse Compton scattering of the soft
  thermal disc photons in a cloud of hot electrons or ``corona''
  (Zdziarski \& Gierlinski 2004), and a reflection component (Ross \&
  Fabian 1993). The latter arises as hard emission from the corona
  irradiates the cooler disc below and results in ``reflection
  signatures'' consisting of fluorescent and recombination emission
  lines as well as absorption features. The most prominent of these
  ``signatures'' is the broad, skewed Fe-\ka\ line observed in a
  number of GBHCs and active galactic nuclei (AGNs, see Miller 2007
  for a recent review) indicative of reflection from the innermost
  regions of an accretion disc.

  In the inner regions of an accretion disc the iron \ka\ line shape
  is distorted by various relativistic effects such as gravitational
  redshift, light-bending, frame-dragging and Doppler shifts, with the
  effects becoming more prominent the closer the line is emitted to
  the event horizon (Fabian et al. 1989, 2000; Laor 1991 ). In the
  case of an accretion disc around a non-spinning Schwarzchild BH,
  stable circular orbits can only extend down to the radius of
  marginal stability, \rms$ = 6.0$\rg where \rg$ = GM/c^{2}$. This
  radius depends on the spin parameter {\it a/M}, and decreases to
  $\approx 1.24$\rg\ for a maximally rotating ($a/M \approx 0.998)$
  Kerr BH (Thorne 1974). By making the standard assumption that the
  emission region extends down to the radius of marginal stability
  (i.e. \rin\ = \rms ) one can obtain an estimate on the dimensionless
  spin parameter (Bardeen et al. 1972; Reynolds \& Fabian 2007).

  GX 339-4 is a dynamically constrained ($M_{BH} \geq 6.0 \msun$;
  Hynes et al. 2003; Munoz-Darias et al. 2008) recurrent black hole
  binary (BHB). Its distance has been estimated at 8\kpc\ (Zdziarski
  et al. 2004). Observations have been made on multiple occasions in
  various spectral states from the ``low/hard'' to the ``very high
  state'' (VHS; for a recent review on the different spectral states
  see e.g. McClintock \& Remillard 2006). In both the VHS and the
  low/hard state (LHS) GX 339-4 shows a power-law spectra, with photon
  index $\Gamma \sim$ 2.5--2.7 and 1.4--1.5 respectively (Miller et
  al. 2004, 2006, hereafter M1 and M2; for a recent analysis of {\it
    Suzaku} observation in the ``intermediate'' state see Miller et
  al. 2008), as well as the presence of a quasi-thermal disc
  component, usually described by a multicolour disc blackbody model
  (MCD; Mitsuda et al. 1984). In both cases, the fluorescent Fe \ka\
  features have been modelled by the addition of a \laor\ relativistic
  line (Laor 1991) plus an ionised disc reflection component (\pexriv
  , Magdziarz \& Zdziarski 1995). In this manner, Miller et al. (2004,
  2006) measured \rin$\sim$ 2.0--3.0\rg\ and \rin$\sim$ 3.0--5.0\rg,
  for the VHS and LHS respectively.

  It has long been known that the reflection in a BHB system cannot be
  mimicked simply by adding a blackbody component to the reflection
  spectrum from an otherwise cool disc. Compton broadening of the iron
  \ka\ line is of greater importance in warm accretion discs and
  should thus modify the spectral behaviour of BHB compared to that of
  AGNs. In this paper we undertake a systematic reanalysis of the VHS
  and LHS of GX 339-4, as reported by Miller et al. (2004, 2006). We
  employ the self-consistent reflection model developed by Ross \&
  Fabian (2007), where blackbody radiation entering the accretion disc
  surface layers from below, as well as the effect of Comptonisation,
  is implicitly included in the model.

  Our method of measuring the spin of stellar mass BH is complementary
  to that of McClintock, Narayan \& Shafee (2007). They use the soft
  high state when any power-law emission is minimal and fit the
  quasi-blackbody continuum spectrum. Their method, in contrast to
  ours, requires accurate measurements of the mass and distance of the
  black hole. In the following section, we detail our analysis
  procedure and results.

\section{Observation and Data reduction}

GX 339-4 was observed in its VHS by \xmm\ for 75.6\ks, starting on
2002 September 29 09:06:42 UT (revolution 514) and simultaneously by
\rxte\ for 9.6\ks\ starting at 09:12:11:28 UT (M1). LHS observations
were made by \xmm\ during revolutions 782 and 783, for a total
exposure of 280\ks\ starting on 2004 March 16 16:23:41 TT and \rxte\
on 2004 March 17 at 12:03:12 TT, observation 90118-01-06-00 (M2). For
the 2002 observation the \epicpn\ camera (Struder et al. 2001) was
operated in ``burst'' mode with a ``thin'' optical blocking
filter. For the low/hard observation the \epicmos1\ and \epicmos2\
cameras (Turner et al. 2001) were operated in the standard
``full-frame'' mode with the ``medium'' {\it EPIC} optical blocking
filter in place. Starting with the unscreened level 1 data files for
all the aforementioned observations, we followed the reduction
procedures mentioned in M1 and M2.

In essence, for the \xmm\ observation of the VHS the Observational
Data Files (ODFs) were processed using the latest \xmm\ {\it Science
  Analysis System \thinspace v 7.1.0 (SAS)}, with events being
extracted in a stripe in RAWX (31.5-40.5) {\it vs} RAWY (2.5-178.5)
space. Bad pixels and events too close to chip edges were ignored by
requiring ``FLAG = 0'' and ``PATTERN $\leq$ 4''. The energy channels
were grouped by a factor of five to create a spectrum. The standard
canned burst mode response file {\it epn\_bu23\_sdY9\_thin.rsp} was
used to fit the spectrum. The total good exposure time selected was
2.2\ks. Due to the high source flux, background spectra were not
extracted. \rxte\ data for the VHS were reduced using the \rxte\
tools provided in the {\it HEASOFT \thinspace \rm v 5.2} software
package. Standard time filtering, including the South Atlantic Anomaly
and elevation angle restriction ( $\geq 10\rm\thinspace degs.$ from
the earth's limb) resulted in a net Proportional Counter Array ({\it
  PCA}) and High-Energy X-Ray Timing Experiment (\hexte) exposures of
9.3 and 3.3\ks, respectively. To account for residual uncertainties in
the calibration of {\it PCU-2}, we added 0.75\thinspace per cent}
systematic error to all its energy channels. The response matrix was
made by the task ``pcarsp''. The \hexte\ source and background spectra
were made using the standard recipes. Standard canned response were
used for spectral fitting.

\begin{figure}

\rotatebox{270}{
\resizebox{!}{7.8cm} 
{\includegraphics{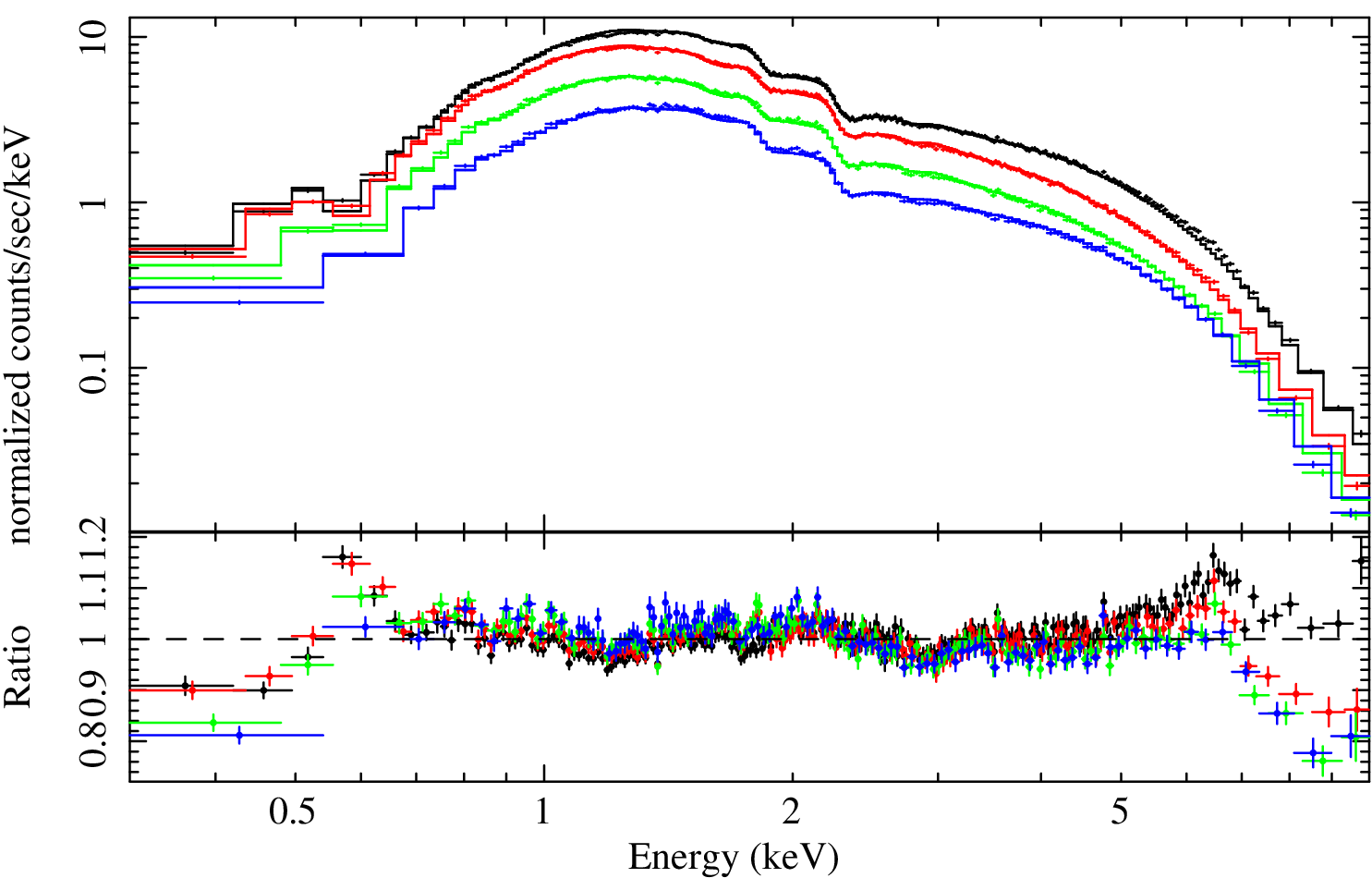}}
}
\end{figure}

\begin{figure}
\
\includegraphics[angle=270,width=7.9cm]{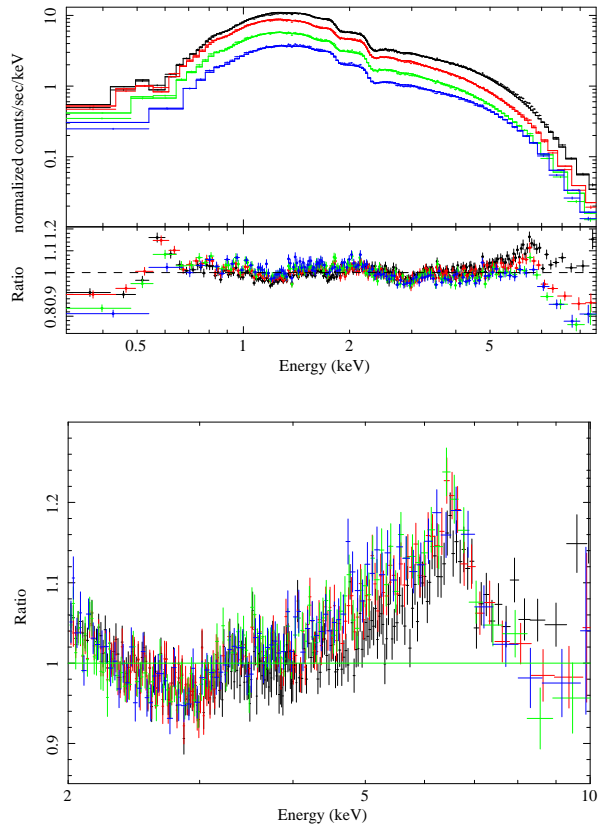}

\caption{\epicmos1, revolution 782 spectra for various source
  extraction regions and event selection criterion fitted with a
  simple powerlaw and MCD component(see text). {\it Top: From top to
    bottom:} Source extraction region in annulus with inner radius
  18'' single-quadruple pixel, 18'' single pixel, 30'' single
 pixel, and 50'' single-quadruple pixel. {\it Bottom:} Data/model
  ratio for the region around the iron line profile with a
  powerlaw. The energy range 4.0-7.0\kev\ was ignored during the
  fit. It is clear that pile-up is only significantly present in the
  annular with inner radius 18'' and both single-quadruple pixel
  ({\it top black}).}

\end{figure}

The \xmm\ data from the 2004 LHS was reduced using {\it SAS\thinspace
  \rm v\thinspace 7.1.0}. As opposed to the \epicpn\ data, the
\epicmos\ cameras in the ``full frame'' mode are more susceptible to
photon and pattern pile-up. Pile-up occurs when several photons hit
two neighbouring (pattern pile-up) or the same (photon pile-up) pixel
in the CCD before the end of a read-out cycle. If this happen the
events are counted as one single event having an energy equal to the
sum of all their energies, thus hardening the spectra. In order to
investigate the effect of pile-up suffered by the \epicmos\ cameras,
we used the SAS tool {\it xmmselect} to obtain spectra from annular
regions of inner radius 18'', 25'', 30'' and 50'' and outer radius of
120'' centered on the source. The events were filtered by requiring
``FLAG=0'' and ``PATTERN $ \leq $ 12'' (single--quadruple pixel
events) as well as ``PATTERN==0'' (only single-pixel
events). Background spectra were extracted from a 60'' circle near the
corner of the central chip of each {\it MOS} camera for both ``PATTERN
$ \leq $12'' and ``PATTERN==0''. Response files for each spectrum were
created using the tools {\it rmfgen} and {\it arfgen}. The {\it FTOOL
  grppha} was used to require at least 20 {\sc counts\thinspace
  bin$^{-1}$}. The spectra of four different extraction region and
event criterion for the \epicmos1\ (revolution 782) observation are
shown in Figure 1, fitted with a simple powerlaw and MCD component
modified by absorption in the interstellar medium ({\rm PHABS} model
in \xspec). The various parameter were tied between the spectra and a
normalisation constant was allowed to float between them. It is clear
from Fig. 1 that pile-up only significantly affects the spectrum
created with the source extraction region with inner radius 18'' and
``PATTERN$\leq$12''. All other spectra are consistent with the most
conservative extraction region (inner radius 50'', single-pixel
events) at energies between 0.7--10.0 \kev. The overall shape of the
spectrum is, however, not significantly affected by pile-up in all
extraction regions and patterns, as can be seen in the lower panel of
Fig.1.

In order to maximise signal-to-noise and make use of the best
calibrated response matrix for the LHS, we use the spectra extracted
from the annulus with inner radius of 18'' and single-pixel events
throughout the analysis detailed in this work. A net exposure time of
56 and 59\ks\ was obtained in revolution 782 for the \epicmos1 and 2
camera respectively and $58\ks$ for each camera in revolution 783. For
the \rxte\ data set, the reduction procedure involved the use of the
tools provided in the \rxte\ {\it HEASOFT \thinspace \rm v 6.0}
software package. We used the ``Standard 2 mode'' data from {\it
  PCU-2} only. The event files and spectra were screened and the
background and response files created. Systematic errors of
0.75{\thinspace per cent} were added to all {\it PCU-2} energy
channels. The \hexte -A cluster was operated in the ``standard archive
mode''. Background-subtracted spectrum and associated instrument
response files were created using standard procedures. The \rxte\
observations resulted in net {\it PCA} and \hexte\ exposures of 2.2
and 0.8\ks, respectively.

We restrict our spectral analyses of the \xmm\ \epicpn\ data to the
0.7--9.0\kev\ band. For a preliminary constraint on the blackbody
temperature of the LHS we use \xmm\ \epicmos\ data in the range
0.5--2.0\kev\ as we expect the temperature to be low, however for the
rest of the analyses \xmm\ \epicmos\ is used in the range
0.7--9.0\kev, similarly to the VHS, unless stated otherwise.  The {\it
  PCU-2} spectrum is restricted to the 2.8-25.0\kev\ band with an edge
at 4.78\kev\ ($\tau=0.1$) to account for the strong Xe {\it L}
edge. \hexte\ spectrum is analysed between 20.0--100.0\kev. A Gaussian
line at 2.31\kev\ is introduced when fitting the \epicpn\ spectrum due
to the presence of a feature at this energy that resembles an emission
line. This feature is likely to be caused by Au M-shell edges and Si
features in the detectors (M1). When fitting the \rxte\ spectra as
well as the four spectra from the \xmm\ low/hard observation, ({\it
  MOS}1 and 2 for revolution 782 and 783), a joint fit is achieved by
allowing a normalisation constant to float between the various
spectra. All parameters in fits involving different instruments were
tied. \xspecv\ (Arnaud 1996) was used to analyse all spectra. The
quoted errors on the derived model parameters correspond to a
90{\thinspace per cent} confidence level for one parameter of interest
($\Delta\chi^{2} = 2.71$ criterion), unless otherwise stated.

\section{analysis and results}

\subsection{Fits to \rxte\ Data:  2.8--100.0\kev\ Continuum  }

We first analyse the \rxte\ {\it PCU-2} and \hexte\ spectra in order
to constrain the power-law continuum of the two states. By considering
the simplest power-law plus MCD model, modified by absorption in the
interstellar medium (\phabs\ model in \xspec) with an equivalent
neutral hydrogen column density fixed at \nh$ = 5.3\times 10^{21}
\cm^{-2}$ (Kong et al. 2000), resulted in a poor fit for both states,
with $\chi^{2}/\nu = 279.5/74$ and 275.0/76 for the VHS and LHS
respectively. In the case of the LHS the addition of a MCD component
did not affect the fit. Significant residual features are present in
the region around the Fe \ka\ fluorescence line. In order to
phenomenologically model a disc reflection line we initially added a
Gaussian emission line and smeared edge components (\smedge\ in
\xspec) to the model. This significantly improved the fit with
$\chi^{2}/\nu = 68.7/68$ and 88.9/70 for the VHS and LHS
respectively. The parameters measured in the VHS for this model are
$\Gamma_{\rm PL}  = 2.56^{+0.12}_{-0.08}$, ${\it R}_{\rm PL} =
2.7^{+1.5}_{-0.5}$, {\it kT} $ = 0.87^{+0.02}_{-0.06}\kev$, ${\it
  R}_{\rm MCD} = 2200^{+1000}_{-300}$, ${\it E}_{\rm gauss}  =
6.0^{+0.4}_{-1.0}\kev$, {\rm FWHM}$ = 2.6^{+0.9}_{-1.0}\kev$, ${\it
  R}_{\rm gauss}  = 0.02^{+0.07}_{-0.01}$, ${\rm EW}  = 220\pm
100\ev$, ${\it E}_{\rm smedge}  = 8.8^{+0.5}_{-1.7}\kev$, $\tau_{\rm
  smedge}  = 0.2^{+0.8}_{-0.2}$, ${\it W}_{\rm smedge}  =
2.0^{+3.0}_{-1.0}\kev$, (where {\it R} is the normalisation for each
function). The equivalent parameters for the LHS are $\Gamma_{\rm PL} 
= 1.48\pm 0.01$, ${\it R}_{\rm PL}  = 0.204^{+0.005}_{-0.007}$, ${\it
  E}_{\rm gauss}  = 5.9^{+1.2}_{-0.6}\kev$, {\rm FWHM} $ =
0.9^{+1.4}_{-0.5}\kev$, ${\it R}_{\rm gauss}  =
1.7^{+2.8}_{-1.0}\times 10^{-3}$, {\rm EW} $ = 110\pm 40\ev$, ${\it
  E}_{\rm smedge}  = 7.1^{+0.6}\kev$, $\tau_{\rm smedge}  = 0.3^{+0.2}_{-0.1}
$, ${\it W}_{\rm smedge}  = 2.3\pm 2.0\kev$. The values obtained
for both states are in agreement with those in M1 and M2.

\subsection{ Fits to \xmm\ \epicpn\ and {\it MOS} data }

\begin{figure}

\rotatebox{270}{
\resizebox{!}{7.8cm} 
{\includegraphics{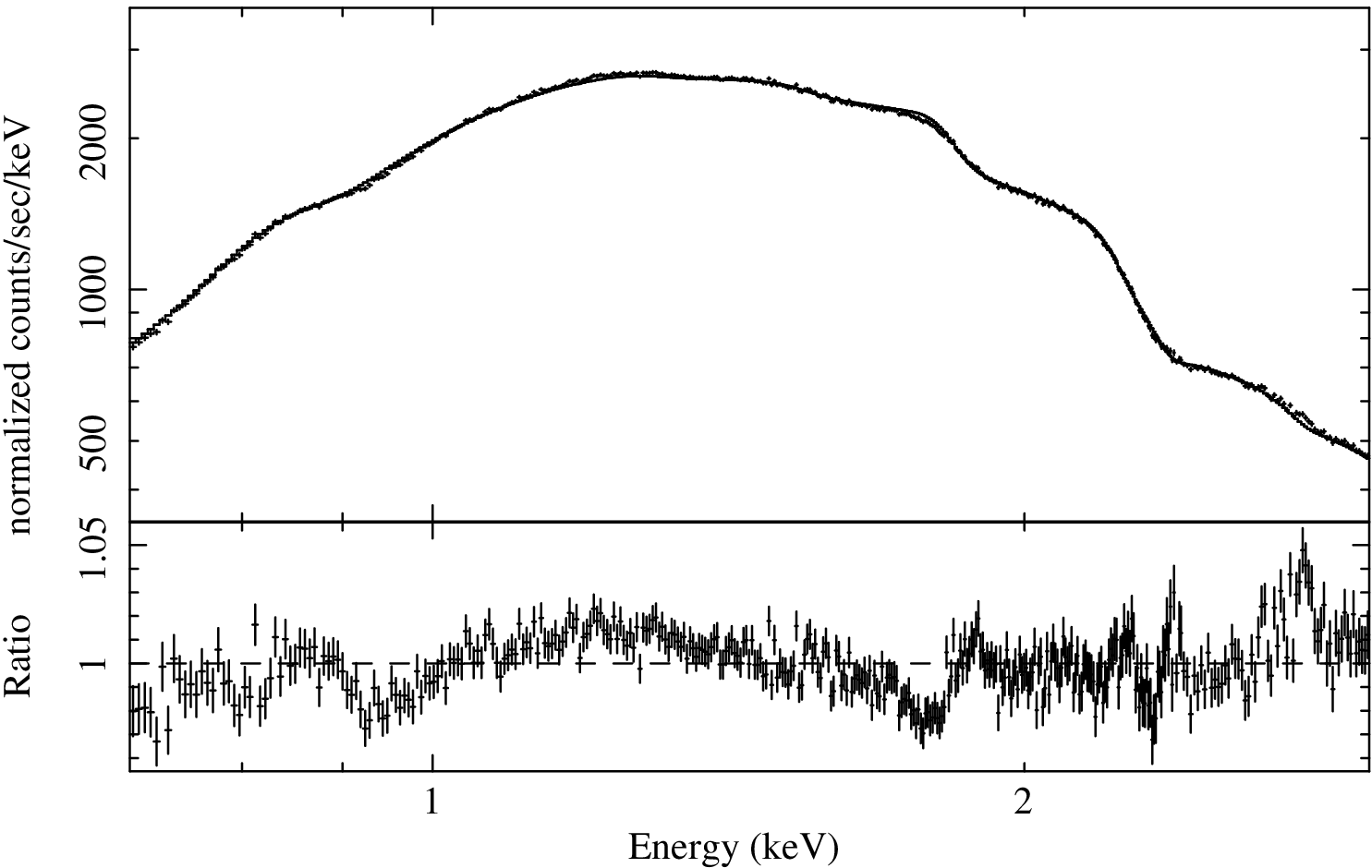}}
}
\rotatebox{270}{
\resizebox{!}{8.cm} 
{\includegraphics{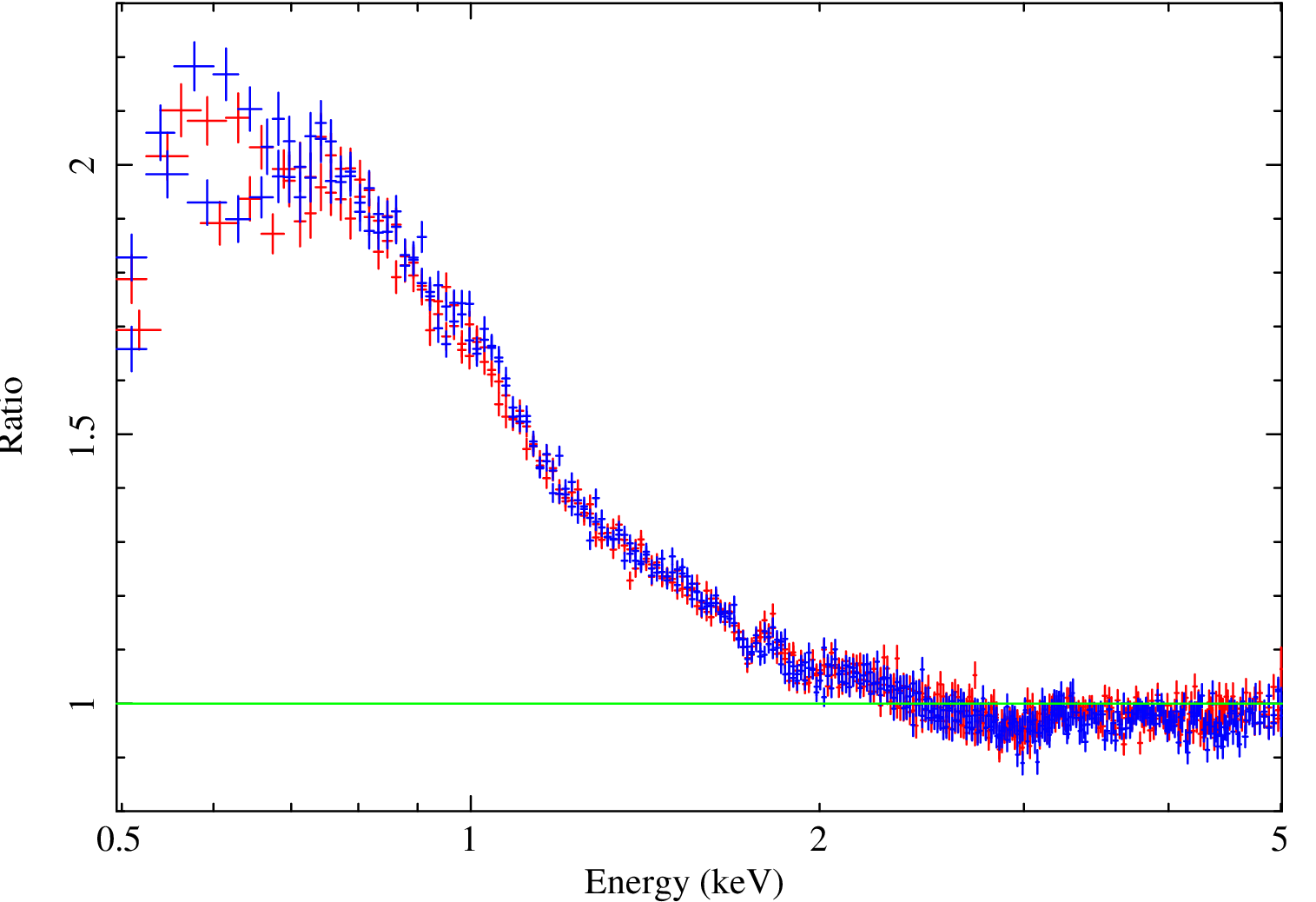}}
}
\caption{{\it Top}: Best fit spectra for the VHS showing the presence of a
  quasi-blackbody with a temperature of $\approx 0.76\kev$ (see
  text). {\it Bottom}: Data/model ratio without a quasi-blackbody component
  for the LHS. Revolution 782 and 783 are shown in red and blue
  respectively (the data are combined for plotting purposes only). It
  is clear from these plots that a semi-blackbody component is present
  in both states.}

\end{figure} 

\subsubsection{Verifying the presence of a quasi-blackbody component}

Whilst fitting \xmm\ data for both VHS and LHS, the power-law index
was constrained to lie within $\Delta \Gamma \leq 0.1$ from the values
obtained in the \rxte\ fits. We began by considering a simple
power-law continuum plus blackbody component in the form of a MCD. The
hydrogen column density was fixed at \nh$ = 5.3 \times 10^{21}
\cm^{-2}$ for both states. A fit to the VHS \epicpn\ data in the range
0.7--3.0\kev\ ($\chi^{2}/\nu = 1448.9/462$) shows the presence of a
blackbody component with temperature of $\approx 0.76\kev$ (Fig. 2,
top). We used \epicmos\ data in the range 0.5--2.0\kev\ to verify the
presence of a MCD component in the LHS. Fig. 2 (bottom) shows the
data/model ratio (extended to an energy range of 5.0\kev) without a
blackbody component for the LHS. The best fit model requires a disc
blackbody with a temperature of $\approx 0.22$, with $\chi^{2}/\nu
=1137.8/389$. The best fit without a quasi-blackbody gives
$\chi^{2}/\nu = 129617.2/389$, which clearly indicates the need for
this component. Recently, similar results have been obtained for the
low/hard state of \gx, where an optically thick disc with a temperature of
$\approx0.2\kev$ has been reported (Tomsick et al. 2008).

\subsubsection{Simple Model: 0.7--9.0\kev}

We fit both the VHS and LHS simultaneously with a power-law plus MCD
component. Only the value of \nh\ was tied between the states. The
best-fit value found for \nh\ of $5.170 \pm 0.001 \times 10^{21}
\cm^{-2}$ is in accordance with that of Kong et al. 2000. We
restricted the value of \nh\ to 5.1--5.3$times 10^{21}$ for the
remainder of thus work. Figure 3 shows the spectra with the data/model
ratio. The formally unacceptable fit ($\chi^{2}/\nu = 10279.4/3860$)
can be seen to be due mainly to the broad iron line and soft energy
residuals, and for the VHS, a large Fe \ka\ absorption edge.

To provide a physically realistic description of the Fe line region we
initially added a relativistic Fe line (\laor, Laor 1991) to the MCD
and power-law continuum and, for the VHS, a smeared edge to
phenomenologically model the iron \ka\ absorption edge. The \laor\
model describes a broad line from an accretion disc surrounding a
rotating Kerr BH, with an emissivity profile described by a power-law
of the form $\epsilon_{(r)} = r^{\it -q}$. The outer disc radius was
fixed at the maximum allowed value of 400\rg. The inner radius of the
disc, \rin, emissivity index, {\it q}, disc inclination, {\it i}, and
the normalization were free to vary. It should be noted that
constraining the spin based on the \laor\ model, although robust, is
only an approximation since the identification of \rin, as determined
from \laor\ assumes a hard wired spin parameter of $a=0.998$. The way
that the inferred BH spin depends on the position of the inner radius
was explored by Dovciak, Karas \& Yaqoob (2004), and was shown to be
consistent with the ``true'' spin as one considers more rapidly
rotating black holes (see their Figure 2).  We fit both the VHS and
LHS individually, restricting the value of \nh\ to $ 5.2 \pm0.1\times
10^{21} \cm^{-2}$. The fit parameters are given in Table 1. Adding
both a \laor\ and \smedge\ components significantly improved the fit
for the VHS, with $\chi^{2}/\nu = 2478.1/1652$ and the LHS with
$\chi^{2}/\nu = 3069.7/2376$ (Fig. 4). For the LHS, residuals below
2.0\kev\ indicates that the simple power-law plus MCD components,
predominant in this range, is not a accurate description of the
continuum, and a more complex reflection component should be
present. If data below 2.0\kev\ is removed for this state, and the
disc temperature, normalisation and column density is frozen, Model 1
converges to $\chi^{2}/\nu = 2249.3/2031$. The energy range below
2.0\kev\ does not affect the direct measurement of the Fe \ka\ line
profile since the \laor\ function used to model the line only extends
down to an energy of $\approx 3.5\kev$ (see Fig. 4). We note that the
parameters values found here differ slightly from those of similar
models in M1 and M2 likely due to the restriction imposed on the
neutral hydrogen column density, \nh\ and on improved calibration.

\subsubsection{More Complex Models: Very High State }

\begin{figure*}
\rotatebox{270}{
\resizebox{!}{12.5cm} 
{\includegraphics{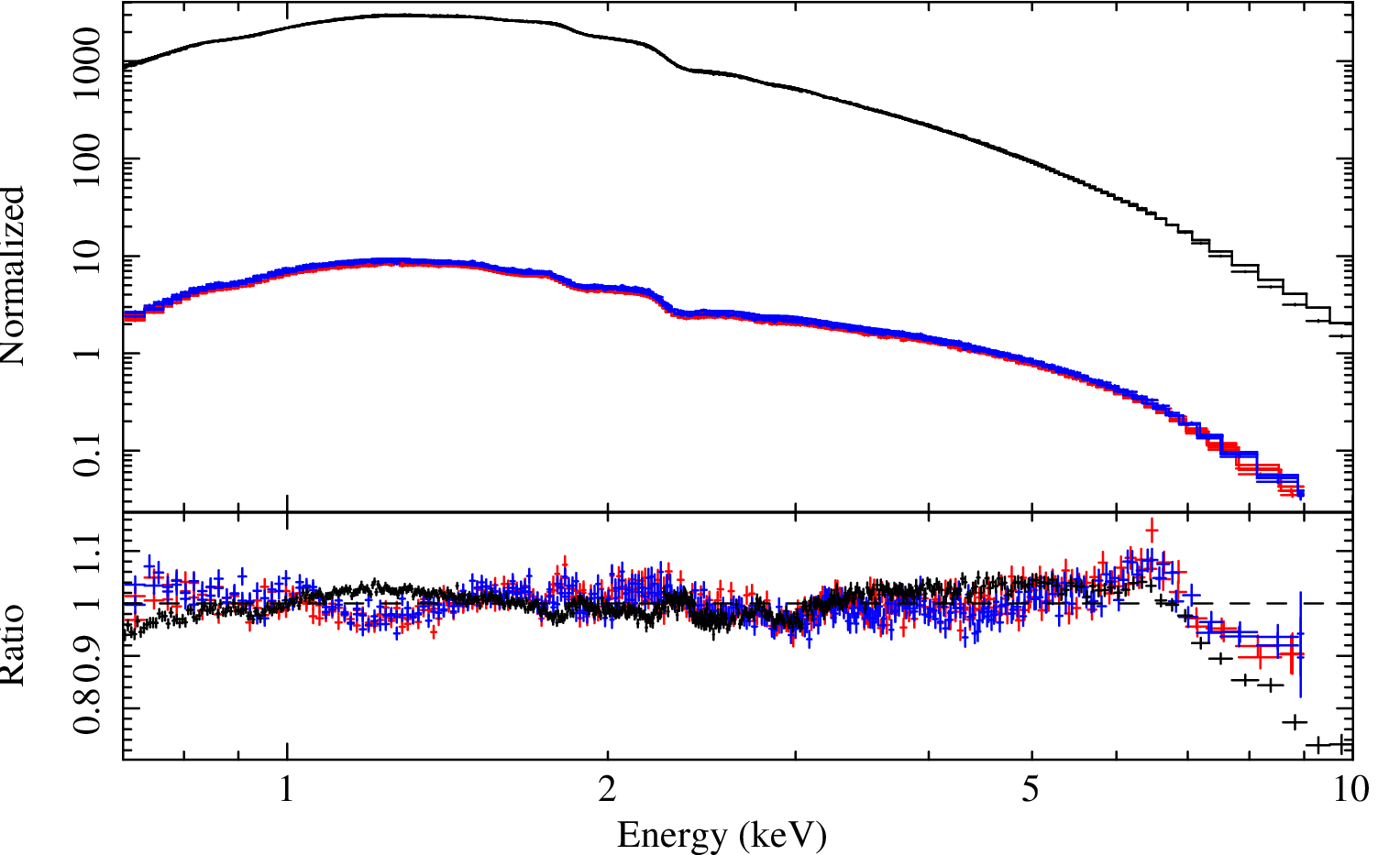}}
}
\end{figure*} 

\begin{figure*}
\rotatebox{270}{
\resizebox{!}{8.5cm} 
{\includegraphics{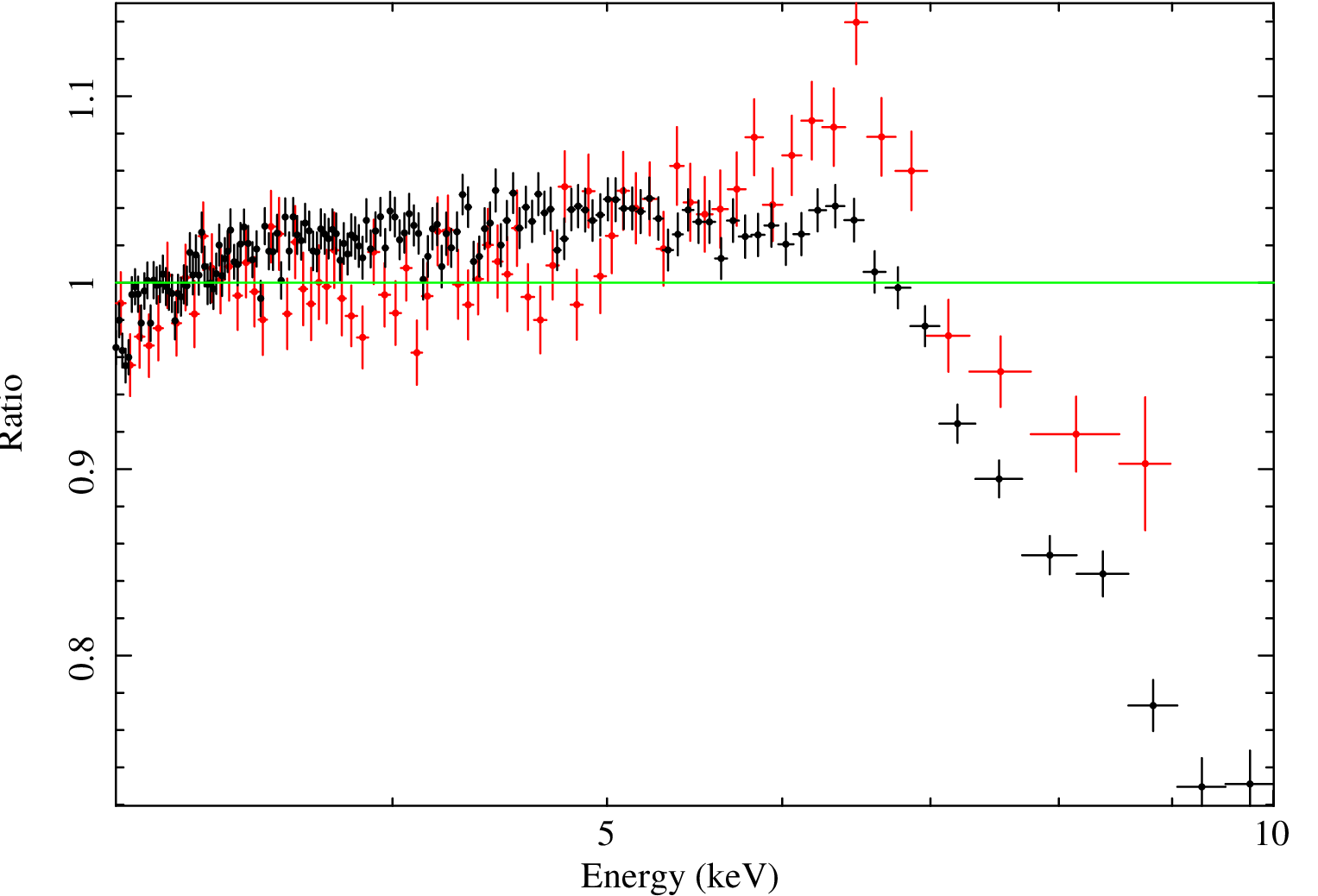}}
}
\caption{{\it Top}: Data/model ratio for a simple model consisting of a
  power-law and MCD components only. The \xmm\ \epicpn\ data for the
  VHS is shown in black. Spectra for the combined \epicmos\ (LHS)
  revolution 782 and 783 are shown in red and blue respectively (the
  data are combined for plotting purposes only). {\it Bottom}: Blowup of the
  \epicpn\ (black) and \epicmos1\ (revolution 782, red) spectrum
  showing the broad iron line and Fe \ka\ edge region. The data have
  been re-binned for visual clarity. The \epicpn\ spectrum has been
  extended to 10\kev\ for illustration purposes only. }

\end{figure*}

\begin{figure*}
\rotatebox{270}{
\resizebox{!}{9cm} 
{\includegraphics{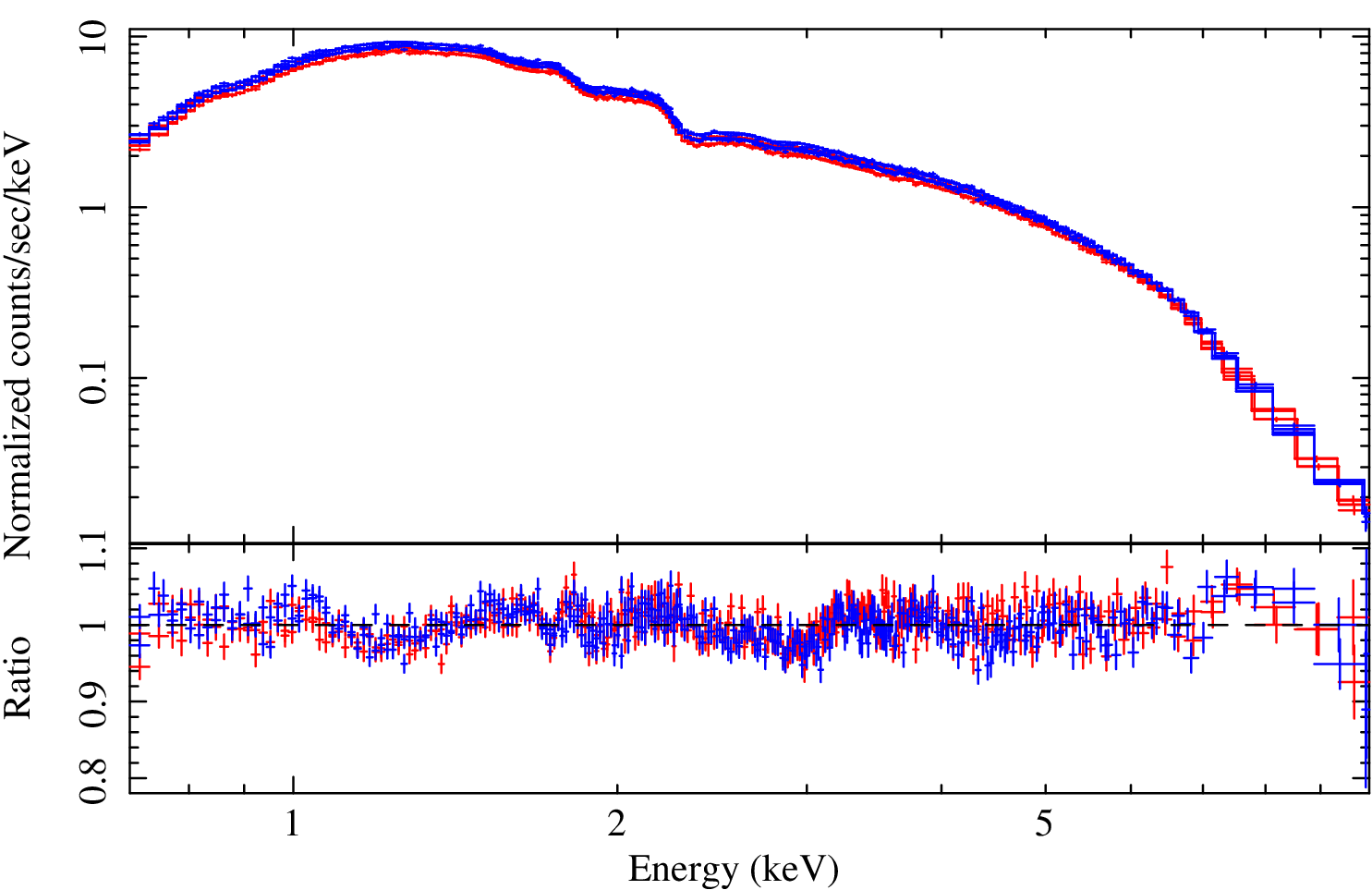}}
}
\rotatebox{270}{
\resizebox{!}{7.5cm} 
{\includegraphics{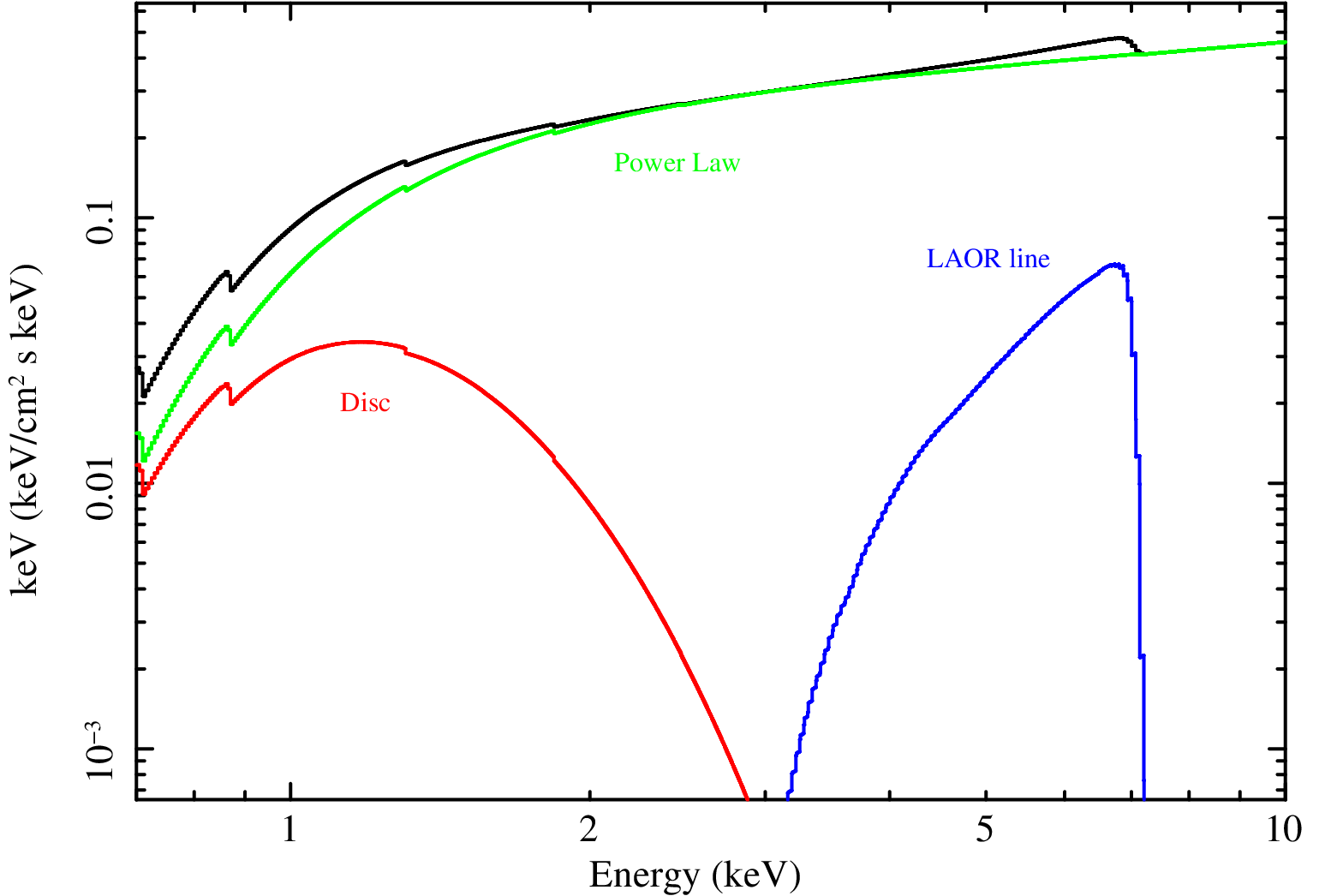}}
}

\caption{{\it Left}: LHS data/model ratio for a model consisting of a simple
  power-law and MCD components, as well as a \laor\ line (Model
  1). Spectra for the combined \epicmos\ revolution 782 and 783 are
  shown in red and blue respectively. The data are combined and
  re-binned for plotting purposes only. {\it Right}: Model components for
  {\it MOS} 1 revolution 782.  }

\end{figure*}

\begin{table}
  \caption{Results of Fits to \xmm\ \epicpn\ (VHS) and \epicmos\ (LHS) data. The spectra were analysed in the 0.7--9.0\kev\  band. }
\begin{center}
\begin{tabular}{lcccccccccc}                
\hline
\hline
\multicolumn{5}{c}{{\small{Model 1 ( PHABS~$\times$~SMEDGE~$\times$~[~PL~+~DISKBB~+~LAOR~])}}} \\
\hline
Parameter & Very High State & Low/Hard state \\
\hline
\nh\ $(10^{21} \cm^{-2})$ & $5.300_{-0.004}$ & $5.100^{+0.006} $\\
$\Gamma$ & $2.567^{+0.004}_{-0.01}$ & $1.700_{-0.002} $\\
$R_{\rm PL}$ & $2.87^{+0.02}_{-0.01}$ & $0.231 \pm 0.005$\\
{\it kT} (\kev) & $0.721\pm0.001$ & $0.235\pm0.02 $\\
$R_{\rm MCD} $ & $2890^{+10}_{-30}$ & $8100\pm300 $\\
$E_{\rm Laor} (keV)$ & $6.97_{-0.01}$ &  $6.97^{+0.01}_{-0.06}$\\
$q_{\rm Laor}$ & $6.82^{+0.03}_{-0.04}$ & $3.23^{+0.05}_{-0.04} $\\
\rin (\rg) & $1.91^{+0.02}_{-0.01}$ & $2.8\pm0.1 $\\
$i ({\rm deg})$ & $18.2^{+0.3}_{-0.5}$ & $10.0^{+2.0} $\\
$R_{\rm Laor} (\times10^{-3})$ & $130 \pm 2$ & $3.75\pm0.15$\\
$E_{\rm smedge} (keV)$ & $7.10 ^{+0.01}$ & ... \\
$\tau_{\rm smedge}$ & $2.3^{+0.1}_{-0.6}$ & ... \\
$W_{\rm smedge}$ & $4.1\pm0.2$ & ... \\
$\chi^{2}/\nu$ & $2478.1/1652$ & $3069.7/2376 $\\
\hline
\hline
\end{tabular}
\end{center}
\end{table}

In all our previous fits, the presence of a broad iron emission line
has been determined and successfully modelled by the \laor\
kernel. The presence of a possible edge at $\approx7.1\kev$ found for
the VHS is consistent with that predicted from absorption due to Fe
{\it K}-shell transition in partially ionised, ``warm'', material
(Ross \&\ Fabian 1993; Ross et al. 1996). To date, BHB spectra have
been modelled by a combination of a line model such as \laor\ or {\sc
  KERRDISK} (Brenneman \&\ Reynolds 2006), a separate reflection
function such as \pexriv\ and a multicolour disc, since no
self-consistent reflection model had been available. Here, we use the
reflection model developed by Ross \& Fabian (2007, \refhiden) to
model those components jointly for the VHS. The parameters of the
model are the number density of hydrogen in the illuminated surface
layer, ${\it H}_{\rm den}$, the value of {\it kT} for the blackbody
entering the surface layer from below, the power-law photon index, and
the ratio of the total flux illuminating the disc to the total
blackbody flux emitted by the disc. The disc reflection spectra is
convolved with relativistic blurring kernel \kdblur, which is derived
from the code by Laor (1991). The power law index of \refhiden\ is
tied to that of the hard component. We constrain the value of \nh,
inclination, and power-law index to 5.1--5.3$\times10^{21}\cm^{-2}$,
10--20 degrees and 2.5--2.7 as found from Model 1 above. Using this
model we obtained a much improved fit for the VHS, with $\chi^{2}/\nu
= 2348.8/1655$ (Model 2). Allowing for a broken power-law for the
emissivity further improves the fit, with $\chi^{2}/\nu = 2237.8/1653$
(Model 3, see Fig. 5 top) and an F-Statistic value over the previous
fit of 41, implying a probability of $< 10^{-18}$ of a random
occurrence. The parameters found for these models are shown in Table
2. A direct inspection of Fig. 5 (Top) shows the presence of a
possible photo-ionisation edge for {\rm O VIII} at $\approx 0.86$\kev\
and a narrow line at $\approx 6.4\kev$. Adding an edge at 0.86\kev\
with an optical depth $\tau = 3.5^{+ 0.7}_{-0.9}\times 10^{-2}$, as
well as a narrow Gaussian at 6.4\kev\ results in $\chi^{2}/\nu =
2144.9/1651$ (Fig. 5 bottom). The presence of the narrow Gaussian
component is required at the 98{\thinspace per cent} level (F-test
probability of 0.02) and can be attributed as due to reflection from
distant materials. By looking at the contribution to $\chi^{2}$ (not
shown in Figure 5) it is clear that the majority of the contribution
comes from the energy range 1.6--3.3\kev\ and is likely the result of
{\rm Au} M-shell edges and {\rm Si} features in the detector. Due to
the high signal--to-noise ratio achieved, these features are clearly
revealed and thus affect the overall $\chi^{2}$ fit statistics but not
the values of the best-fit parameters.

\begin{table}
\begin{center}
  \caption{Result of more complex fits to \xmm\ \epicpn\ data for \gx\
    in the Very High State.}

\begin{tabular}{lcccccccccc}                
\hline
\hline
Parameter & Model 2 & Model 3 \\
\hline
\nh\ $(10^{21} \cm^{-2})$ & $5.16\pm0.02$ & $5.17\pm0.03 $ \\
$\Gamma$ & $2.7_{-0.01}$ & $2.7_{-0.02} $ \\
$R_{\rm PL}$ & $1.45^{+0.25}_{-0.15}$ & $2.8^{+0.2}_{-0.3}$  \\
{\it kT} (\kev) & $0.519^{+0.006}_{-0.01}$ & $0.554\pm 0.004$ \\
$H_{\rm den}(\times10^{21}{\thinspace {\rm H}\cm^{-3})} $ & $4.05\pm0.2$ & $4.52^{+0.4}_{-0.2} $ \\
{\it Illum/BB} & $4.4^{+0.4}_{-0.3}$ &  $2.0\pm0.1$ \\
$R_{\refhiden} $ & $4.4^{+0.2}_{-0.1}$ & $2.7^{+0.1}_{-0.3}$ \\
$q_{\rm in}$ & $6.84^{+0.1}_{-0.2}$ & $7.6^{+0.3}_{-0.6} $ \\
$q_{\rm out}$ & $...$ & $3.7^{+0.3}_{-0.8}$ \\
$r_{\rm break} (r_{g})$ & $...$ & $4.9^{+0.6}_{-0.7} $ \\
\rin (\rg) & $1.804^{+0.08}_{-0.004}$ & $2.03^{+0.025}_{-0.035} $ \\
{\it i} (\rm deg) & $19.5^{+0.5}_{-3}$ & $19.98^{+0.02}_{-0.7} $ \\
$\chi^{2}/\nu$ & $2348.8/1655$ & $2237.8/1653 $ \\
\hline
\hline
\end{tabular}
\end{center}

\small Notes.- Model 2 is described in \xspec\ as (PHABS~$\times$~(~KDBLUR~$\times$~[PL~+~REFHIDEN~])). Model 3 assumes the accretion disc has a broken-power law emissivity profile described by the function KDBLUR2 in \xspec. The value of  \nh, inclination, and power-law index was constrained to 5.1--5.3$\times10^{21}\cm^{-2}$, 10--20 degrees and 2.5--2.7, respectively in both models. 

\end{table}

\begin{figure}
\rotatebox{270}{
\resizebox{!}{8.cm} 
{\includegraphics{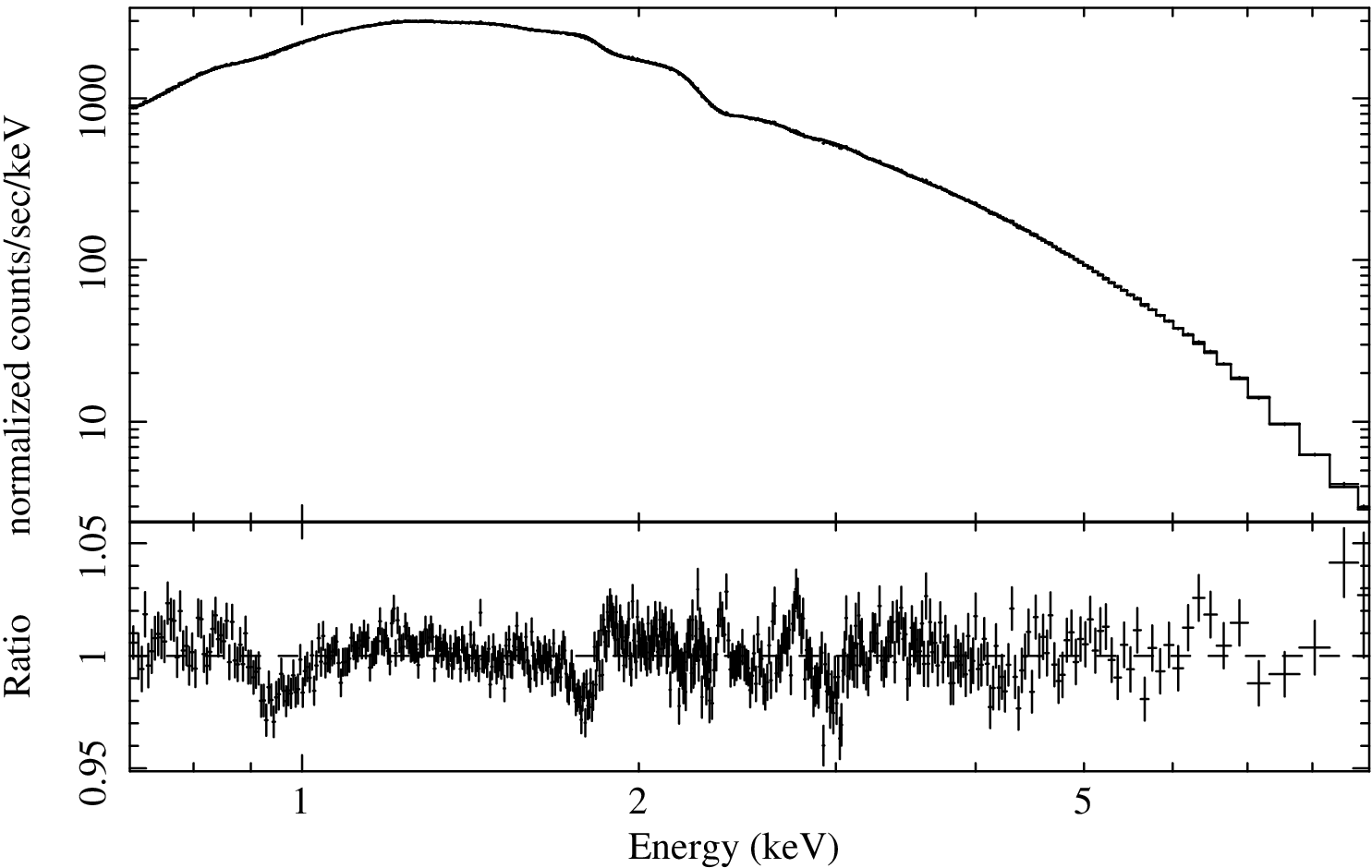}}
}
\rotatebox{270}{
\resizebox{!}{8.cm} 
{\includegraphics{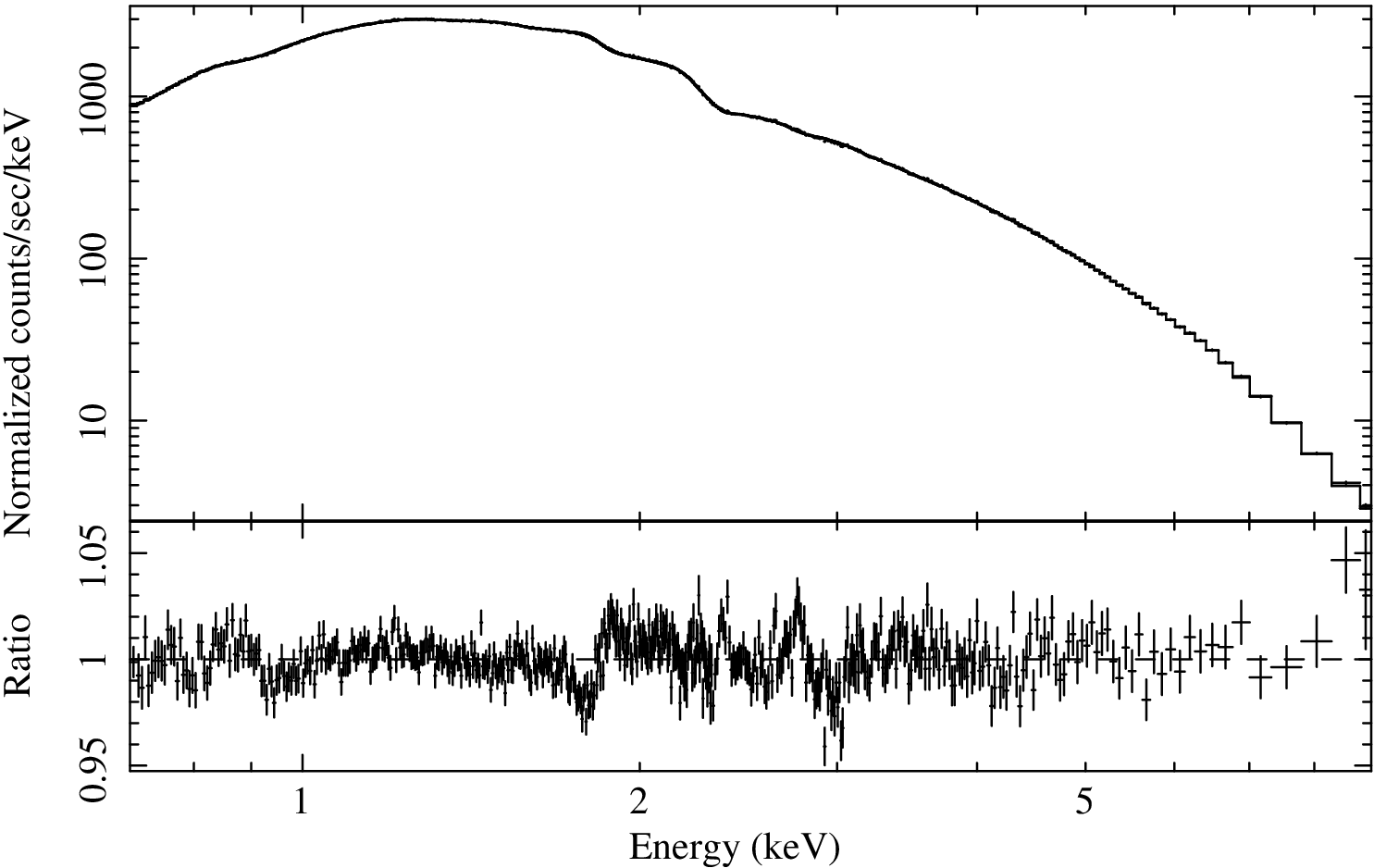}}
}

\caption{Data/model ratio for the VHS. {\it Top}: Model assumes a
  broken power-law emissivity profile and constitute of a power-law
  and the disc reflection function \refhiden. The presence of a {\rm O
    VIII} edge at 0.86\kev and a possible narrow emission line at
  $\approx6.4\kev$ can be seen. {\it Bottom}: Same as above but with
  an additional narrow Gaussian line and a {\rm O VIII} edge at
  0.86\kev. The data have been re-binned for visual clarity. }
\end{figure} 

\begin{figure}
\rotatebox{270}{
\resizebox{!}{9cm} 
{\includegraphics{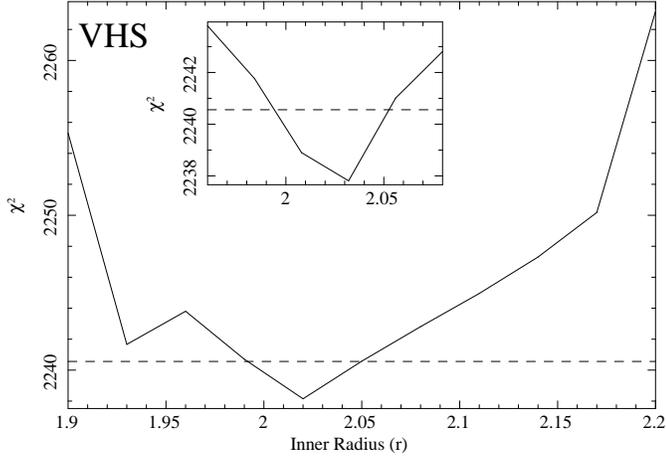}}
}
\caption{ $\chi^{2}$ vs \rin\ plot for \gx\ in its Very High State. A
  value of \rin$ = 2.03^{+0.025}_{-0.035}$\rg\ is found at the 90
  per cent confidence level for one parameter, $\Delta\chi^{2} = 2.71$
  criterion shown by the dashed line. The inset shows a close up of
  the region around the $\chi^{2}$ minima.}
\end{figure}

From the irradiating source luminosity, $L_{\rm x}$ (defined here
between 0.2--10.0 \kev), disc hydrogen-density ($H_{\rm den}$) and
radius, {\it R} (approximated to 2.0\rin), it is possible to estimate
the ionisation parameter, ($\xi = L_{\rm x}/H_{\rm den} R^{2}$), of
the VHS as ${\log \xi} \approx 4.2$, which is consistent with previous
values found for \gx\ in its very high state (M1). Using a
self-consistent reflection model we were able to constrain the value
of the innermost radius to \rin = $2.03^{+0.025}_{-0.035}$\rg. This
strong constraint can be better appreciated in Fig. 6, where the
90{\thinspace per cent} confidence lever for \rin\ is shown as the
dashed-line in the $\chi^{2}$ plot, obtained with the ``steppar''
command in \xspec. Assuming that \rin\ = \rms\ we get a spin parameter
{\it a/M}$= 0.939^{+0.004}_{-0.003}$ for the VHS.

It should be noted that the measured values for the disc temperature
found by the \refhiden\ model is at least 0.15\kev\ less than the
value obtained via fits with the additive components (Model 1). This
behaviour is as expected (see Merloni, Fabian \& Ross 2000; Ross \&
Fabian 2007) since in Model 1 the soft disc component is modelled with
a multi-colour disc quasi-blackbody ({\sc DISKBB}) having a
Comptonised surface colour-temperature ($T_{col}$) intrinsically
higher than the effective mid-plane temperature ($T_{eff}$) used in
\refhiden.  The spectral hardening factor $f_{col}=T_{col}/T_{eff}$
(Shimura \& Takahara 1995) of $\approx 1.3$ found for the VHS is very
similar to the conventional value of $1.7 \pm 0.2$ (Shimura \&
Takahara 1995). It is expected that a similar reduction in the value
of {\it kT} should occur for the LHS where the effective disc
temperature would be less than the value found in section 3.3.1 of
$\approx 0.22\kev$.

\subsubsection{More Complex Models: Low/Hard State }

\begin{figure}
\rotatebox{270}{
\resizebox{!}{8.5cm} 
{\includegraphics{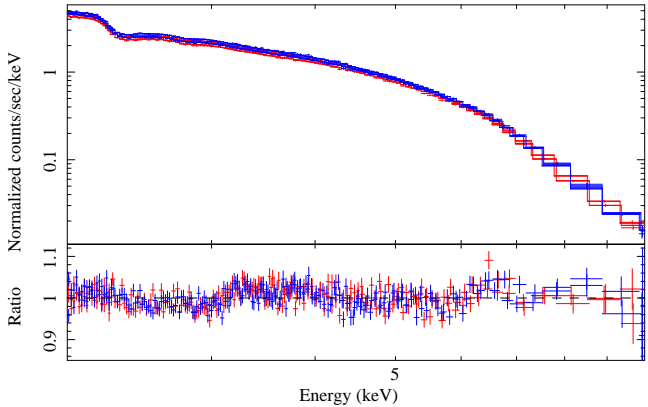}}
}
\caption{LHS data/model ratio for a relativistic blurred disc
  reflection model (\reflionx) and power-law as described in the text. The spectra for the combined \epicmos\
  revolution 782 and 783 are shown in red and blue respectively. The
  data have been combined and re-binned for plotting purposes only.}
\end{figure}

\begin{figure}
\rotatebox{270}{
\resizebox{!}{9.2cm} 
{\includegraphics{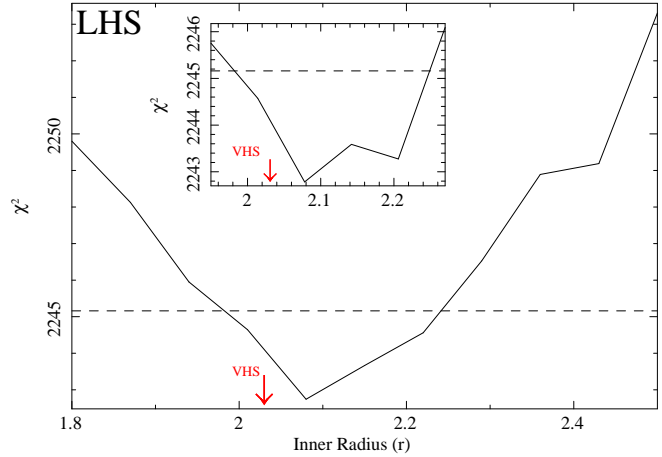}}
}
\caption{$\chi^{2}$ vs \rin\ plot for \gx\ in its Low/Hard State. An
  inner radius, \rin$ = 2.08^{+0.17}_{-0.10}$\rg\ is found at the
  90{\thinspace per cent} confidence level for one parameter,
  $\Delta\chi^{2} = 2.71$ criterion shown by the dashed line. The
  inset shows how $\chi^{2}$ behaves closer to the minima. The inner radius found for the VHS is shown in red for comparison. }
\end{figure}

The effective temperature of around 0.15\kev\ and hydrogen number
density ${\it H}_{\rm den} > 2\times10^{21} {\rm H\thinspace cm^{-3}}$
expected for the LHS falls outside the parameter range of the
\refhiden\ model. However, the model is being developed and an
analyses of the LHS with \refhiden\ is left for future work. The
disc-blackbody in the LHS has a negligible effect on the iron-\ka\
features above 2.0\kev (see Figure 4). For comparison with our results
for the VHS, we use \reflionx\ to analyse \xmm\ \epicmos\ data for the
LHS in the range 2.0--10.0\kev\ with a model similar to \refhiden\ but
lacking the intrinsic blackbody disc component. The model \reflionx\
is a revised version of {\sc REFLION} (Ross \& Fabian 2005) used to
describe reflection from accretion disc in AGN systems where the
blackbody emission is at too low an energy to affect the Fe K$\alpha$
emission. It should be stressed that the reflection features above
2\kev\ are unlikely to be significantly affected by the change from
\refhiden\ to \reflionx. The parameters of the model are the iron
abundance (set to solar), photon index of the illuminating power-law,
ionisation parameter, $\xi$, and the normalization. The disc
reflection spectra is convolved with the relativistic blurring kernel,
\kdblur. The power law index in \reflionx\ is tied to that of the hard
component. We constrain the value of the inclination, and power-law
index to 10--20 degrees and 1.4--1.6, respectively. The hydrogen
column density, \nh\ is fixed at $5.17 \times10^{21}\cm^{-2}$, the
best fit value found for the VHS, as we do not expect it to vary. The
best fit obtained with the blurred \reflionx\ model is shown in Fig.7
and detailed in Table 3. This model gives $\chi^{2}/\nu = 2242.5/2031$
with an emissivity index of $3.065\pm 0.05$ indicating a standard
``lamp-post'' emissivity profile. The value found for the inclination
of $20_{-1.7}${\rm \thinspace deg.} is in agreement with that for the
VHS. The low disc ionisation parameter, log($\xi$)$\approx 3$, is
consistent with that expected for low disc temperatures. At 90 per
cent confidence, this model gives constraint on the innermost stable
radius of \rin$=2.08^{+0.17}_{-0.10}$\rg\ (see Fig. 8). If we include
the energy band 0.7-2.0\kev\ to the above fit, a large low energy
residual is present due to the disc emission. By modeling this with a
\diskbb\ component, a best fit of $\chi^{2}/\nu = 3070.9/2388$ is
achieved in the full 0.7--10.0\kev\ range with a disc temperature of
$0.201\pm 0.003$\kev\ as in \S3.2.1.

\begin{table}
  \caption{Results of fits to \xmm\ \epicmos\ data for the Low/Hard state of \gx. The spectra was analysed in the 2.0--10.0\kev\  band. \nh\ was fixed at the quoted value. }

\begin{center}
\begin{tabular}{lcccccccccc}                
\hline
\hline
\hline
Parameter & Low/Hard State  \\
\hline
\nh\ $(10^{21} \cm^{-2})$ & 5.17 \\
$\Gamma$ & $1.48^{+0.09}_{-0.08}$ \\
$R_{\rm PL}$ &$0.11\pm0.01$\\
{\it q} & $3.065\pm 0.05$ \\
\rin (\rg) & $2.08^{+0.17}_{-0.10}$ \\
{\it i} (\rm deg) & $20_{-1.7}$ \\
$\xi ({\rm erg}\cm\s^{-1})$ & $1350\pm100$ \\
$R_{\rm REFLIONX}(\times 10^{-6}) $ & $4.0^{+0.4}_{-1.0}$ \\
$\chi^{2}/\nu$ & $ 2242.5/2031$\\
\hline
\hline
\end{tabular}
\end{center}

\end{table}

\begin{table}
\begin{center}
  \caption{Joint \xmm\ and \rxte\ spectral fits with relativistic blurred disc reflection models.}

\begin{tabular}{lcccccccccc}                
  \hline
  \hline
  Parameter & Very High State & Low/Hard State \\
  \hline
  \nh $(10^{21} \cm^{-2})$ & $5.100^{+0.004}$ & $5.17$ \\
  $\Gamma$ & $2.583\pm0.007$ & $1.43^{+0.005}_{-0.02} $ \\
  $R_{\rm PL}$ & $2.61^{+0.13}_{-0.09}$ & $0.094^{+0.005}_{-0.004}$  \\
  $kT (\kev)$ & $0.585\pm0.001$ & ... \\
  $H_{\rm den}(\times10^{21}{\thinspace {\rm H}\cm^{-3})}$ & $6.6\pm0.2$ & ... \\
  $Illum/BB$ & $1.00\pm0.02$ &  ... \\
  $\xi ({\rm erg}\cm\s^{-1})$ & $>10000$ & $1330^{+70}_{-60} $ \\
  $R_{\refhiden} $ & $1.92^{+0.02}_{-0.06}$ & ... \\
  $R_{\reflionx} (10^{-6}) $ & ... & $4.4\pm0.2$ \\
  $q_{\rm in}$ & $7.05^{+0.05}_{-0.20}$ & $3.16 \pm 0.05$ \\
  $q_{\rm out}$ & $3.0^{+0.1}$ & ...  \\
  $r_{\rm break} (r_{g})$ & $6.0^{+0.2}_{-0.1}$ & ... \\
  \rin (\rg) & $2.02^{+0.02}_{-0.06}$ & $2.04^{+0.07}_{-0.02}$ \\
  {\it i} (\rm deg) & $20.0_{-0.3}$ & $20.0_{-1.3}$  \\
  $\chi^{2}/\nu$ & $2549.3/1718$ & $2316.6/2095$ \\
   \hline
   \hline
\end{tabular}
\end{center}

\small Notes.- VHS was modelled with \refhiden\ and a broken power-law emissivity profile, as described in the text. For the LHS, the disc reflection model \reflionx\ was used, and the spectra was fitted in the range 2--100\kev. 

\end{table}

\begin{figure*}
\rotatebox{270}{
\resizebox{!}{8.5cm} 
{\includegraphics{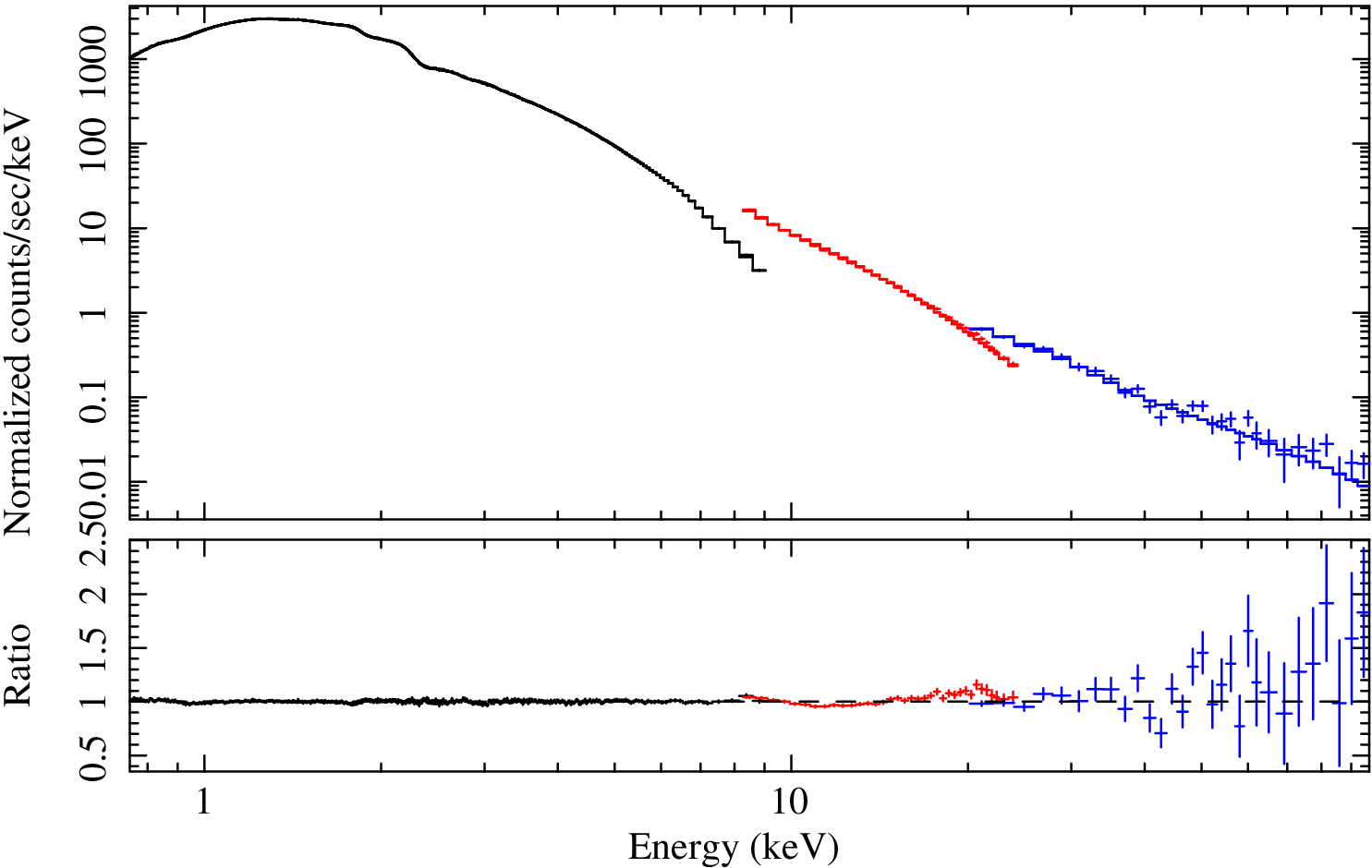}}
}
\rotatebox{270}{
\resizebox{!}{8.5cm} 
{\includegraphics{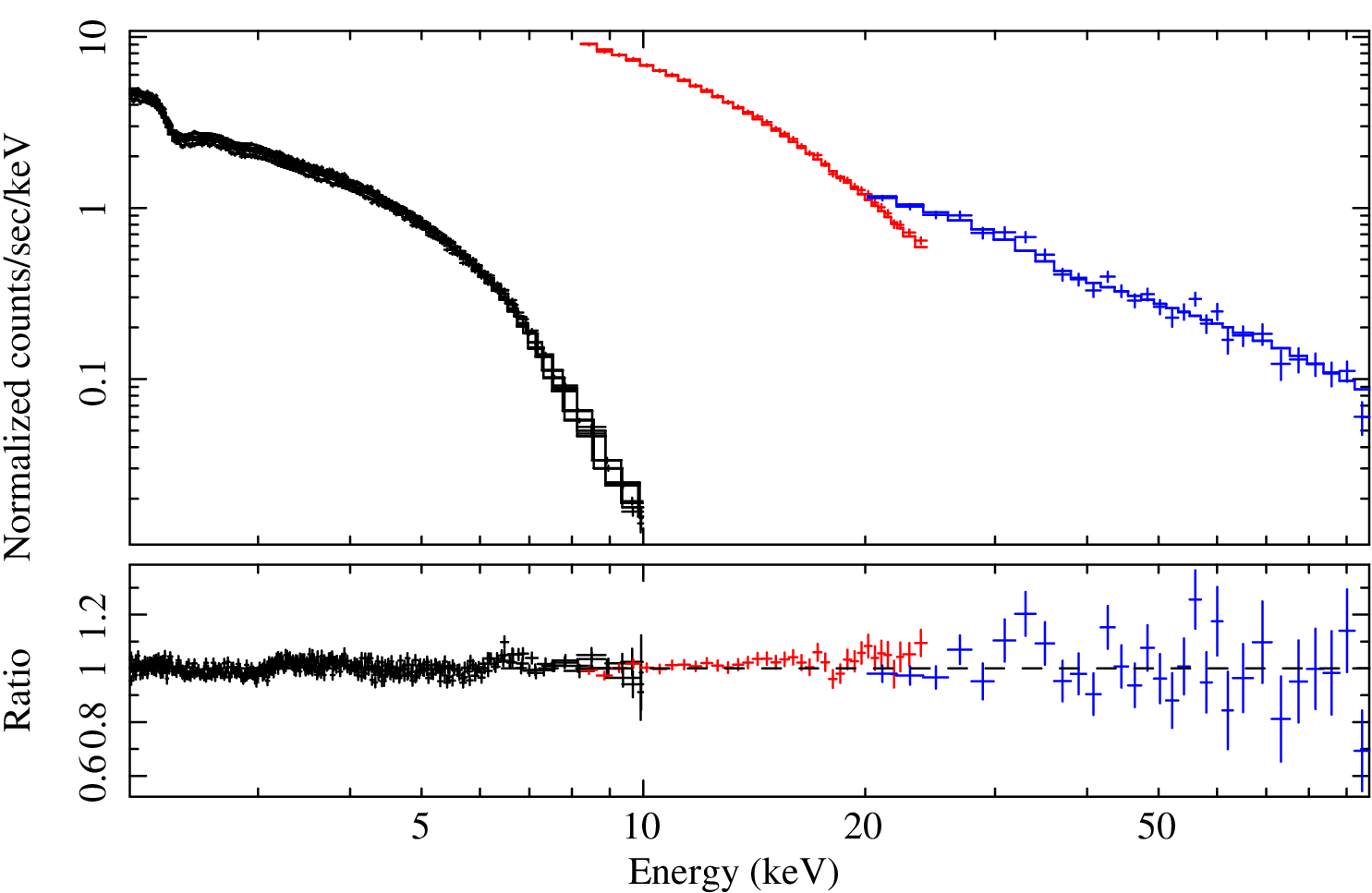}}
}

\caption{\xmm\ and \rxte\ spectra of \gx\, fit jointly with a disc
  reflection model convolved with the relativistic blurring kernel,
  \kdblur. {\it Left}: VHS spectrum in the 0.7--100.0\kev\ range. {\it
    Right}: LHS spectrum in the 2.0--100.0\kev\ range.The best fit
  indicates a value for \rin\ of $2.02^{+0.02}_{-0.06}$\rg\ and
  $2.04^{+0.07}_{-0.02}$\rg\ for the VHS and LHS
  respectively. \epicmos /pn data are shown in black. \rxte\ PCU-2 and
  \hexte\ data are shown in red and blue respectively. The data have
  been re-binned for visual clarity. }
\end{figure*}

\subsection{Joint \xmm\ -- \rxte\ spectrum analysis}
\subsubsection{Very High State: 0.7--100.0\kev}

In order to check the robustness of our results, we extended the fit
from the \epicpn\ spectrum to include the energy range 0.7--100\kev,
using \rxte\ data. PCU-2 data was fitted between 8.0--25.0\kev. This
resulted in a poor fit, with $\chi^{2}/\nu = 4017.0/1727$. It should
be stressed that the quality of the \epicpn\ data far out-weights that
of \rxte\ and thus a statistically worst fit is inevitable in the full
range. However, most of the residuals are accounted for by allowing
the power-law index to vary. The best fit value of
$\Gamma=2.583\pm0.007$ is in accordance to that found in section 3.1
for the \rxte\ continuum. The parameters for fits to the combined
\epicpn\ and \rxte\ spectrum are listed in Table 4 and shown in Fig. 9
(left). It is clear that the model is a very good description of the
spectrum (see Fig. 9), with $\chi^{2}/\nu = 2549.3.0/1718$ in the full
0.7--100.0\kev\ range.  Most importantly, the value for the inner
radius, \rin\ found here of $2.02^{+0.02}_{-0.06}$\rg\ is similar to
that found in section 3.2.3.

\subsubsection{Low/Hard State: 2--100\kev}

A similar extension on the energy range of the LHS was made, with
\rxte\ data being used in conjunction with \epicmos. A best fit of
$\chi^{2}/\nu = 2316.6/2095$ over the full 2.0--100.0\kev\ energy band
is found. The various parameters are shown in Table 4 and the
resulting spectra in Fig. 9 (right). The value for the inner
radius, \rin\ found here of $2.04^{+0.07}_{-0.02}$\rg\ is similar to
that found in section 3.2.4.

\section{Discussion}

\begin{figure*}
  \rotatebox{270}{ \resizebox{!}{8.cm}
    {\includegraphics{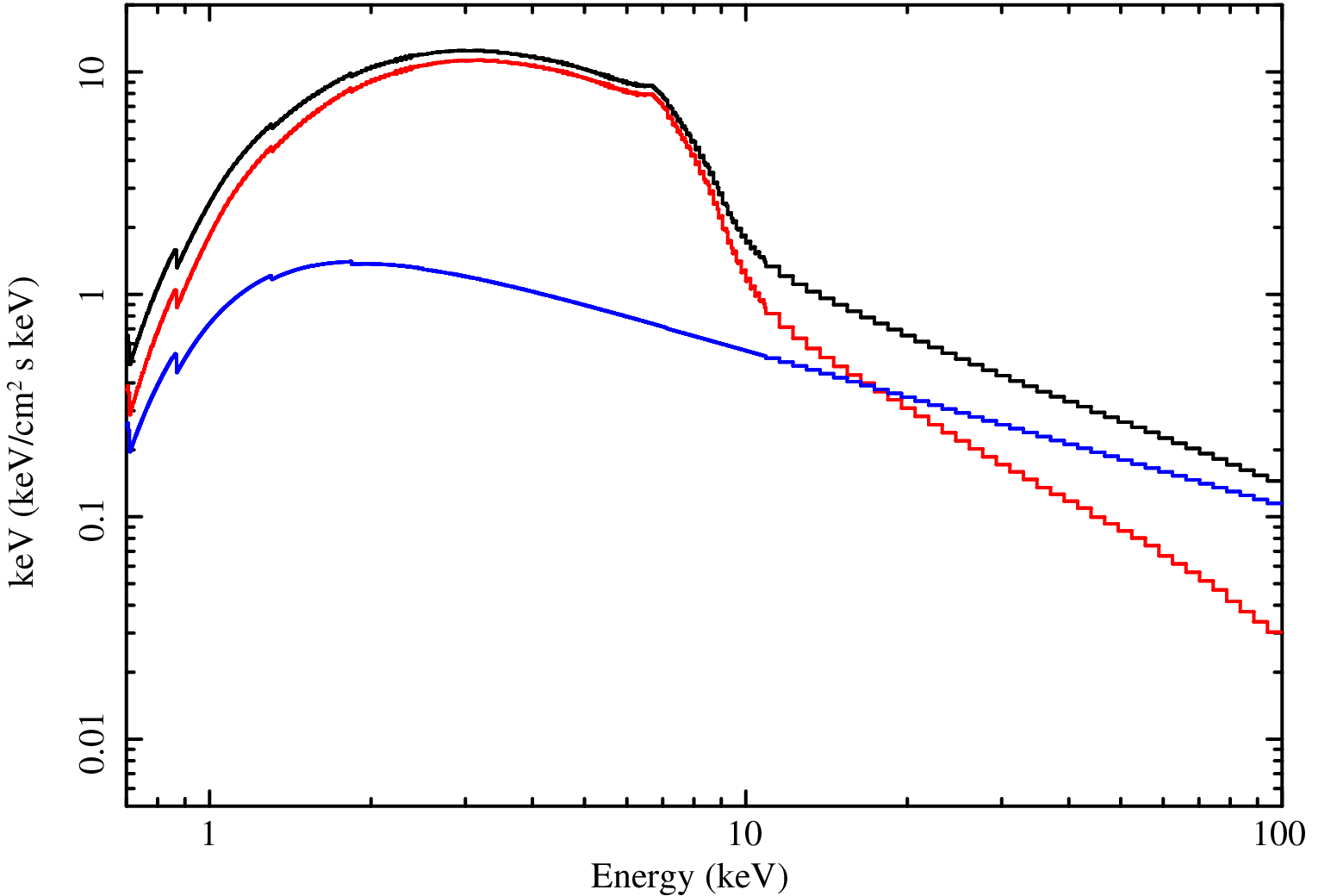}} 
}
 \rotatebox{270}{
 \resizebox{!}{8.cm} 
{\includegraphics{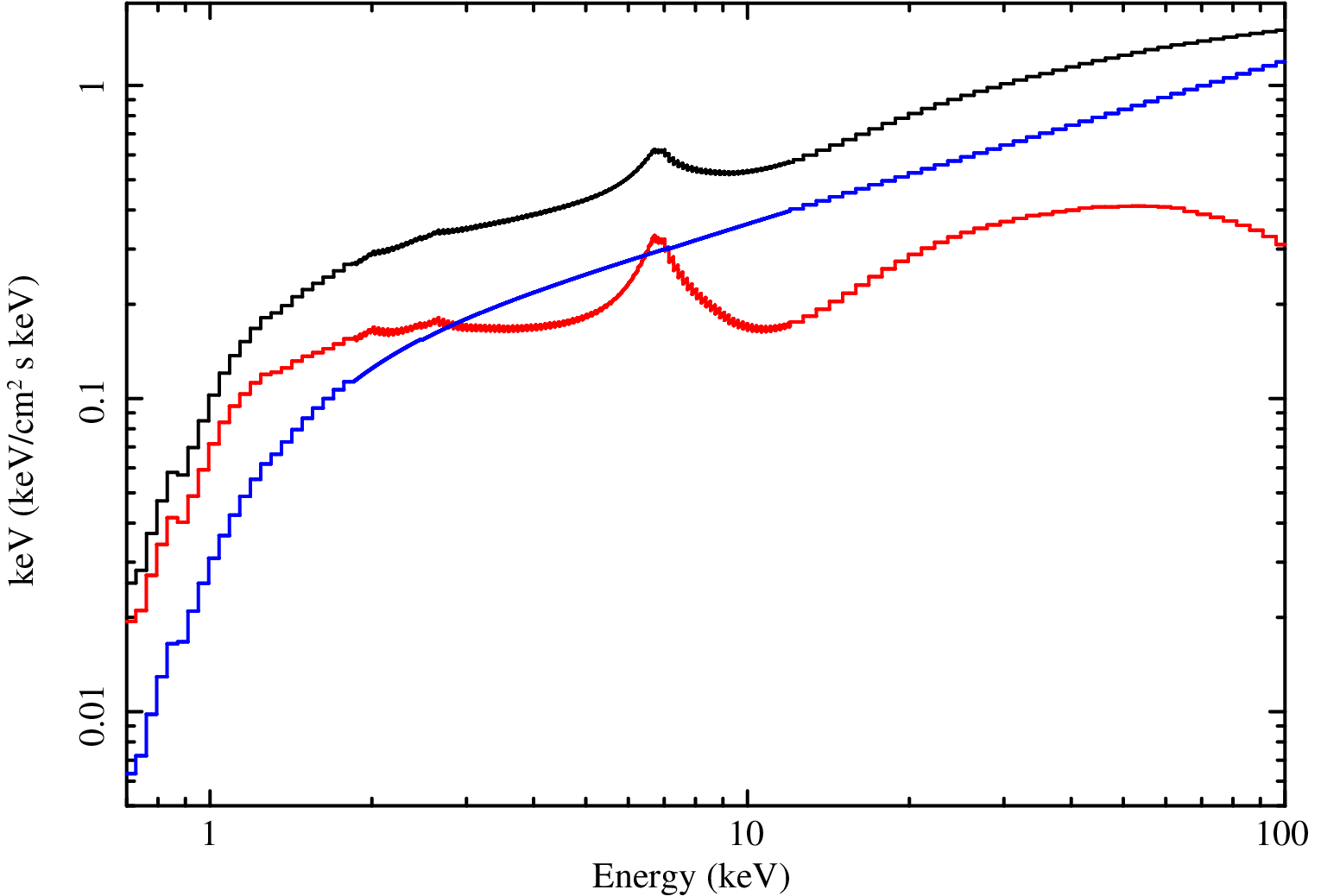}} }
\end{figure*}
\begin{figure*}
\rotatebox{270}{
 \resizebox{!}{8.cm} 
{\includegraphics{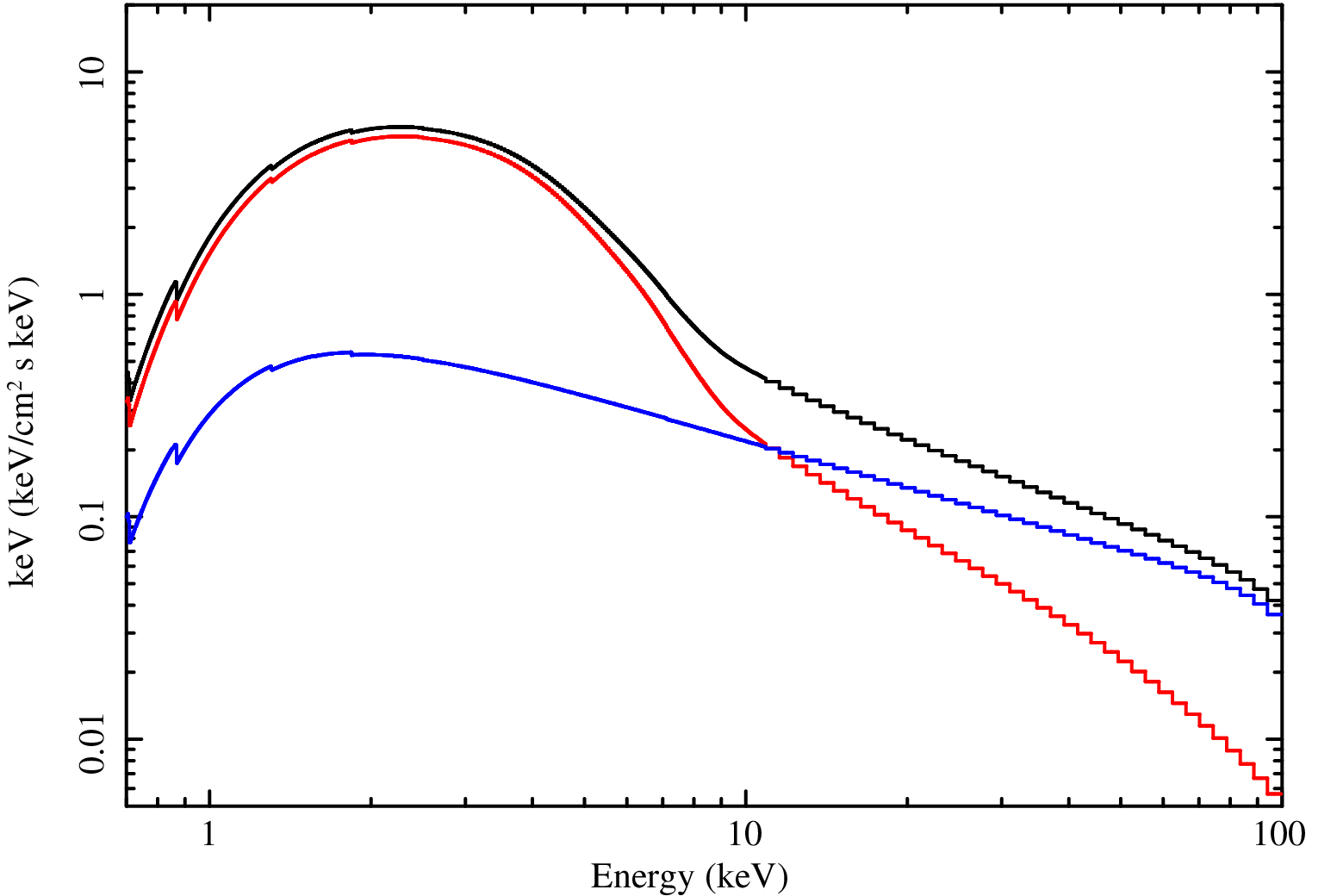}} }
\rotatebox{270}{
 \resizebox{!}{8.cm} 
{\includegraphics{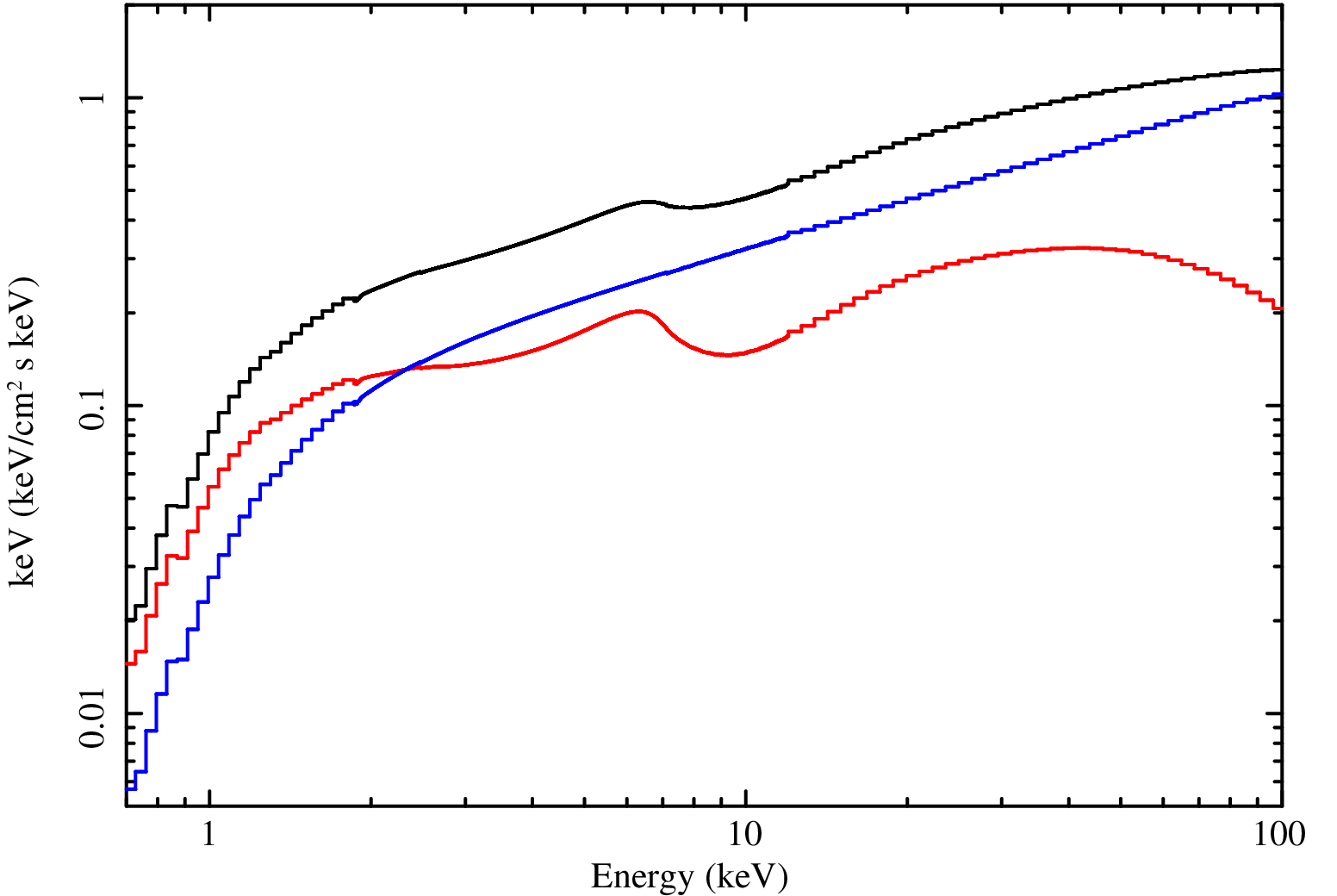}} }

\caption{{\it Top}: Best-fit model components prior to blurring for
  the VHS ({\it left}) and LHS ({\it right}). The combined model,
  reflection, and power-law components are shown in black, red  and
  blue respectively. Iron is fully ionised in the VHS, thus resulting
  in a very small Fe-\ka\ line emission and a large iron-K absorption
  edge. In the LHS, moderate ionisation means that the contribution to
  the spectra from Fe-\ka\ line emission is greater (see text).  For
  the LHS, we only show \epicmos1\ revolution 782 for visual
  clarity. {\it Bottom}: Same as above, with relativistic blurring.}
\end{figure*}

The spectral modelling of the VHS suggests that the surface layer of
the accretion disc is highly ionised, with $\xi \sim 10^4 {\thinspace
  \rm erg}\cm\s^{-1}$. In this state, the iron in the outer layer of
the disc is fully ionised and regions $\gtrsim2$ Thomson mean free
paths below the surface produces a strong iron-K absorption
edge. Narrow \ka\ line emission from this region is then
Compton-broadened as it scatters out of the disc. The strong presence
of the iron-K edge in the VHS can be seen in Fig. 3 and quantitatively
appreciated by the high optical depth ($\tau = 2.3^{+0.1}_{-0.6}$) of
the (phenomenological) component \smedge\ in Model 1. The best-fit
\refhiden\ model clearly shows the large K-shell absorption feature
and weak \ka\ emission line characteristic of the VHS in \gx\
(Fig. 10, left). Note that similar features have also been observed in
the VHS of Cygnus X-1 (Done et al. 1992). The
ionisation parameter found for the LHS, $\xi= 1350\pm100{\thinspace \rm
  erg}\cm\s^{-1}$, is consistent with the disc being moderately
ionised and having a low apparent temperature.  In this state, the
illuminated accretion disc results in a strong Fe-\ka\ line emission
from the top layers. Compton-broadening, although present, cannot
explain the highly broadened and skewed line shape (see Fig. 3), where
the low energy red wing extend down to $\approx 4\kev$. Figure 10
(right) shows the best-fit \reflionx\ model, prior to (top) and after
blurring (bottom), for the LHS. As opposed to the VHS, in the low/hard
state the Fe-\ka\ line is clearly seen. The value for the ionisation
parameter for the low/hard state of $\xi \sim 10^4 {\thinspace \rm
  erg}\cm\s^{-1}$ reported by Tomsick et al. (2008) is an order
of magnitude higher than the present result. At these values, the iron
is fully ionised and should not produce an iron \ka\ reflection line
(Matt, Fabian \& Ross 1993; Young, Ross \& Fabian 1998). The apparent
inconsistency in their results can be attributed to the use of {\sc
  PEXRIV} (Magdziarz \& Zdziarski 1995) as the reflection model. This
model does not account for diffusion of photons in the disc and thus
rectify broadening caused by Comptonisation by increasing the
ionisation parameter.

The obvious differences in the resulting spectra of the two states can
be ascribed to the different ionisation states of the disc. Previous
attempts to model the spectra of Galactic BHB used the relativistic
blurring of the \ka\ line to obtain the innermost radius, \rin\ and
thus the spin parameter. In the present work, the full reflection
spectra for the two extreme states was convolved, and the blurring
parameters were obtained not just from the \ka\ line but from all of
the reflection features. This is particularly important in the case of
the VHS, where the Fe-\ka\ emission line is not the dominant feature
of the reflection component and Compton scattering needs to be fully
accounted in the reflection model. In this state, a steep inner disc
emissivity index of $ q_{\rm in}= 7.6 ^{+0.3}_{-0.6}$, within a radius
$r_{\rm break}= 4.9 ^{+0.6}_{-0.7}$\rg\ is required, indicating that
the corona is centrally concentrated. The model constrains the inner
radius of the accretion disc to \rin$ = 2.03^{+0.025}_{-0.035}$\rg\ at
the 90{\thinspace per cent} confidence level. It should be noted that
the value for \rin\ quoted above for the \xmm\ observation is
consistent with that found for the full full \xmm\ plus \rxte\ fits,
indicating that the model is an accurate description of both the
reflection features as well as the underlying continuum. Assuming that
emission within the innermost stable orbit is negligible (see Reynolds
\& Fabian 2007), the value of \rin\ found here translates to a black
hole spin of $0.939^{+0.004}_{-0.003}$ for the VHS. In the LHS,
spectral fitting using the model \reflionx\ resulted in an inner
radius of \rin$ = 2.08^{+0.17}_{-0.10}$\rg. This translates to a spin
parameter of $0.93^{+0.015}_{-0.02}$ for the LHS.

The value for the spin parameter found for both states of \gx\ are
within one per cent of one another and falls within one sigma
error. It has been argued that bleeding of the iron line emission to
regions inside the innermost stable radius may cause systematic errors
in the derived value of the spin parameter (Reynolds \& Begelman 1997;
Krolik 1999). Using a high-resolution 3-D MHD simulation of a
geometrically-thin accretion disc, Reynolds \& Fabian (2007) have
shown that the ionisation edge is within $\sim0.5$\rg\ of the
innermost stable circular orbit for a non-spinning BH. However, it was
shown by the same authors that this bleeding decreases as the position
of the innermost radius approaches the event horizon, and hence the
spin inferred from the position of \rin\ becomes much closer to the
true spin as one considers more rapidly rotating black holes. Similar
results were reported by Dovciak, Karas \& Yaqoob (2004, see their
Figure 2). In order to verify our results against any systematic
variation, we modelled the VHS with a {\sc KERRDISK} line profile
(Brenneman \& Reynolds 2006). The spin, which is a free parameter of
the model, was found to be $0.952^{+0.005}_{-0.001}$, lying within
$\approx1${\thinspace per cent} of the value inferred from the
innermost radius.

The reflection model, \refhiden, assumes a single-temperature
accretion disc. Although this is not a realistic claim, we believe it
unlikely to have any significant effect on the inferred
innermost radius of emission. In order to verify this hypothesis, we
approximated a ``real'' disc with inner radius increasing
logarithmically from 2\rg\ to 6.78\rg. Each point on the disc was
assumed to radiate like a blackbody with an effective temperature that
scales with radius as $r^{-3/4}$ starting at 0.9\kev. Within this
region, the illuminating flux scaled as $r^{-6}$. Using the \epicpn\
response file, we modelled 1\ks\ of simulated data with a single temperature
\refhiden. As expected, the model constrained the various parameter,
with an effective temperature of $\approx 0.52\kev$ and an inner
radius of $2.041^{+0.004}_{-0.020}$ \rg. As a further check on any
inconsistency that may arise from using a single temperature disc
reflection model to constrain the spin of the black hole, we
investigated the VHS with a different thermal model, ({\sc
  KERRBB}, Li et al. 2005), which includes relativistic smearing in a
disc with radial temperature gradient. The black hole spin, a free
parameter in the {\it KERRBB} model, was found to be $0.93\pm 0.02$,
consistent with that inferred from the single temperature \refhiden\
model.

\section{Conclusions}

We have analysed \xmm\ spectra of \gx\ in both its very high and
low/hard states.  Looking at the difference in the spin parameter
between the two states, as well as that derived from the various
independent models for the VHS, we can estimate the systematic error
in the iron line method to be about 1 per cent. By using a reflection
model which intrinsically accounts for Comptonisation and blackbody
emission, we infer that the spin parameter of \gx\ is $0.935 \pm \thinspace 0.01$
(statistical) $\pm \thinspace 0.01$ (systematic).

\section*{Acknowledgements}

RCR and GM acknowledges STFC for financial support. ACF and RRR thanks
the Royal Society and the College of the Holy Cross respectively. CSR
thanks the US National Science Foundation for support under grant
AST~06-07428.


\begin{thebibliography}{}
       
\bibitem{}
Arnaud K.A., 1996,  ASPC, 101, 17A

\bibitem{}
Bardeen J.M., Press W.H. \& Teukolsky S.A., 1972, ApJ, 178, 347

\bibitem{}
Brenneman L.W., Reynolds C.S., 2006, ApJ, 652, 1028B

\bibitem{}
Done C., Mulchaey J.S., Mushotzky R.F., Arnaud K.A., 1992, ApJ, 395, 275

\bibitem{}
Dovciak M. Karas V. \& Yaqoob T., 2004, ApJ, 153, 205 

\bibitem{}
Fabian A.C., Rees M.J., Stella L., White N.E., 1989, MNRAS, 238, 729

\bibitem{}
Fabian A.C., Iwasawa K., Reynolds C.S., Young A.J., 2000, PASP, 112, 1145

\bibitem{}
Hynes R.I., Steeghs D., Casares J., Charles P.A., \& O'Brian K., 2003, ApJ, 583, L95

\bibitem{}
Kong A.K.H., Kuulkers E., Charles P.A., \&\ Homer L., 2000, MNRAS, 312, L49

\bibitem{}
Krolik J.H., 1999, ApJ, 515, L73

\bibitem{}
Laor A., 1991, ApJ, 376, 90

\bibitem{}
 Li, L.-X., Zimmerman, E.R., Narayan, R., McClintock, J.E., 2005, ApJS, 157, 335

\bibitem{}
Magdziarz P., \& Zdziarski A.A., 1995, MNRAS, 273, 837

\bibitem{}
Matt G,. Fabian A.C., \& Ross R.R., 1993, MNRAS, 264, 839

\bibitem{}
McClintock J.E., \& Remillard R.A. 2006 Black hole binaries (Compact stellar X-ray sources), 157-213

\bibitem{}
McClintock J.E., Narayan R., Shafee R., 2007 arXiv:0707.4492v1 [astro-ph]

\bibitem{}
Merloni A., Fabian A.C. \& Ross R.R., 2000, MNRAS, 313, 193

\bibitem{M1}
Miller J.M., Fabian A.C., Reynolds C.S., Nowak M.A., Homan J., et al. 2004, ApJ, 606, L131

\bibitem{M2}
Miller J.M., Homan J., Steeghs D., Rupen M., Hunstead R.W., Wijnands R., Charles P.A., \& Fabian A.C., 2006, ApJ, 653, 525

\bibitem{}
Miller J.M., Reynolds C.S., Fabian A.C., Cackett E.M., Miniutti G., Raymond J. Steeghs D., Reis R.C., Homan J.,2008, arXiv:0802.3882v1 [astro-ph]

\bibitem{}
Miller J.M., 2007, ARAA, 45, 441-79

\bibitem{}
Mitsuda K.,  Inoue H., Koyama K., Makishima K. Matsuoka M., Ogawara Y., Suzuki K., Shibazaki N., \& Hirano T., 1984, PASJ, 36, 741

\bibitem{}
Munos-Darias T., Casares J., Martinez-Pais I.G., 2008, MNRAS, arXiv:0801.3268v1 [astro-ph]

\bibitem{}
Reynolds C.S., Begelman M.C., 1997, ApJ, 448, 109

\bibitem{}
Reynolds C.S., Fabian A.C., 2007, arXiv:0711.4158v1 [astro-ph]

\bibitem{}
Ross R.R., \& Fabian A.C., 1993, MNRAS, 261, 74

\bibitem{}
Ross R.R., Fabian A.C., Brandt W.N., 1996, MNRAS, 278, 1082

\bibitem{}
Ross R.R., \& Fabian A.C., 2005, MNRAS, 358, 211

\bibitem{}
Ross R.R., \& Fabian A.C., 2007, MNRAS, 381, 1697

\bibitem{}
Shimura T., Takahara F., 1995, ApJ, 445, 780

\bibitem{}
Struder L., et al., 2001, A\&A, 365, L18

\bibitem{}
Tomsick A.J., Kalemci E., Kaaret P., Markoff S., Corbel S., Migliari S., Fender R., Bailyn C.D., Buxton M.M., 2008,	arXiv:0802.3357v1 [astro-ph]

\bibitem{}
Thorne K. S., 1974, ApJ, 191, 507

\bibitem{}
Turner M.J.L., et al., 2001, A\&A, 365, L27

\bibitem{}
Young A.J., Ross R.R., \& Fabian A.C., 1998, MNRAS, 300, L11

\bibitem{} 
Zdziarski A.A. \& Gierlinski M., 2004, Prog. Theor. Phys. Suppl. 155, 99

\bibitem{} 
Zdziarski A.A., Gierlinski M., Mikolanjewska J., Wardzinski G., Smith D.M., Harmon A., \&\ Kitamoto S., 2004, MNRAS, 351, 791

\end{thebibliography}
\end{document}